\documentclass[preprint,12pt]{elsarticle}

\let\today\relax
\makeatletter
\def\ps@pprintTitle{%
    \let\@oddhead\@empty
    \let\@evenhead\@empty
    \def\@oddfoot{\footnotesize\itshape
         {} \hfill\today}%
    \let\@evenfoot\@oddfoot
    }
\makeatother

\usepackage{amssymb}
\usepackage{multirow}
\usepackage{subfigure}
\usepackage{color,soul}
\usepackage{footmisc}
\usepackage{adjustbox}
\usepackage{comment}
\usepackage{placeins}
\usepackage[normalem]{ulem}

\useunder{\uline}{\ul}{}

\begin{document}

\begin{frontmatter}



\title{Micro and macro facial expressions by driven animations in realistic Virtual
Humans}



\author{Rubens Halbig Montanha}
\author{Giovana Nascimento Raupp}
\author{Ana Carolina Policarpo Schmitt}
\author{Victor Flávio de Andrade Araujo}
\author{Soraia Raupp Musse}

\affiliation[**]{organization={Virtual Humans Laboratory (VHLab), School of Tecnology},
            addressline={\\Pontifical Catholic University of Rio Grande do Sul, \\Av. Ipiranga, 6681}, 
            city={Porto Alegre},
            postcode={90619-900}, 
            state={RS},
            country={Brazil}}

\begin{abstract}

Computer Graphics (CG) advancements have allowed the creation of more realistic Virtual Humans (VH) through modern techniques for animating the VH body and face, thereby affecting perception. From traditional methods, including blend shapes, to driven animations using facial and body tracking, these advancements can potentially enhance the perception of comfort and realism in relation to VHs. Previously, Psychology studied facial movements in humans, with some works separating expressions into macro and micro expressions. Also, some previous CG studies have analyzed how macro and micro expressions are perceived, replicating psychology studies in VHs, encompassing studies with realistic and cartoon VHs, and exploring different VH technologies. However, instead of using facial tracking animation methods, these previous studies animated the VHs using blendshapes interpolation. To understand how the facial tracking technique alters the perception of VHs, this paper extends the study to macro and micro expressions, employing two datasets to transfer real facial expressions to VHs and analyze how their expressions are perceived. Our findings suggest that transferring facial expressions from real actors to VHs significantly diminishes the accuracy of emotion perception compared to VH facial animations created by artists. 

\end{abstract}




\end{frontmatter}


\section{Introduction}
\label{sec:introduction}
From techniques of body scanning~\cite{magnenat2003automatic} to the highest fidelity and state-of-art Virtual Humans (VH)~\cite{higgins2021ascending}, the methods of modeling and animation advanced, making it possible to create more realistic VHs.  Focusing on VHs expressions, the use of blendshapes~\cite{lewis2014practice}, applied in games and other applications, associated with the Facial Action Coding System (FACs)~\cite{ekman1969nonverbal} allows reproduce the real human emotions in VH~\cite{higgins2023investigating}~\footnote{Draft version made for arXiv: https://arxiv.org/}. According to FACS authors Ekman and Friesen, the definition of each emotion is made based on a set of Action Units (AUs), which consist of specific parts of faces~\cite{friesen1983emfacs,EkmanFACS}. In the same work that proposes the FACs, the authors presented a study about macro and micro facial expressions~\cite{ekman1969nonverbal}. The macro expressions are generally obvious to be perceived and during between 0.5 and 4 seconds. This kind of facial expression typically matches the expression with the context. Contrarily, micro expressions are often misunderstood or not perceived, and during, in general, less than 500 milliseconds and are expressed unconsciously. Ekman~\cite{Ekman2013AnAF} classified the macro and micro expressions into six emotions: anger, fear, happiness, sadness, surprise, and disgust.

Concerning VHs emotions perception, some studies~\cite{araujo2021perceived,katsyri2015review,mori2012uncanny} suggest people are more comfortable with VHs expressing human emotions and reproducing realistic human behaviors. For example, in the work of Tinwell et al.~\cite{tinwell2011facial}, results provided evidence that the perception of uncanny (or feelings of discomfort) in VHs that display inappropriate facial animation can be influenced by the type of emotion the VH is portraying.  M{\"a}k{\"a}r{\"a}inen et al.~\cite{makarainen2014exaggerating} suggested that the intensity of a facial emotion can influence feelings of discomfort.

One technique hugely used to animate a VH face is performance-driven animation using facial tracking techniques~\cite{barros2019face,liu2016real}. These methods involve transferring the real human's activated AUs to the relative blendshape in the VH using a camera for facial tracking. However, during the transfer process, the VH expressions may be smoother when compared with the real human expression~\cite{peres2023can}. Another challenge in VH modeling and animation, not restricted to facial tracking methods, is the Uncanny Valley (UV). According to Mori theory~\cite{mori2012uncanny}, during an interaction or when a VH is observed, the human response may fall into a valley of discomfort conditional on how similar the VH is to a human. Originally, the UV theory ~\cite{mori2012uncanny} was studied and applied to robots; Nowadays, some authors~\cite{araujo2021perceived,macdorman2016reducing,araujo2023evaluating,katsyri2019virtual} also study the impact of UV in virtual humans. 

Still about visual perception area, Zell et al.~\cite{zell2019perception} present how virtual humans are perceived by explaining a general perception model. This general perception model uses an object observed by the user and used as a visual input, being processed by the brain following the bottom-up and top-down processes. All this process is based on a person's life experience, and at the end, the person recognizes the object. This model used to recognize an object can also be applied in the perception of VHs~\cite{zell2019perception}. Understanding how the 3D depth and shapes are perceived in a 2D screen reveals the importance of pre-existing schemes in the design of new VHs, and how people can attach personalities and attributes based on their judgments and the character's appearance.

Previous literature has researched how individuals perceive macro and micro expressions in VHs using modeled animations. For example, Tastemirova et al.~\cite{tastemirova2022microexpressions} investigated the extent to which micro expressions of happiness and anger with different levels of intensities in realistic VHs could influence perceptions of affection, sincerity, and trustworthiness. Hou et al.~\cite{hou2022micro} examined the perception of micro expressions in VHs with different visual styles. The results showed that realistic VHs can be perceived as having significantly more intense emotions than stylized VHs. Queiroz et al.~\cite{queiroz2014investigating} investigated whether people's perception of emotional expressions in realistic VHs was similar to that in real humans. Following the same methodology used in Queiroz et al. work, Andreotti and collaborators~\cite{andreotti2021perception} studied the user's perception of macro and micro expressions using a cartoon character, i.e., studying the impact of a non-realistic VH. Recently, Montanha et al.~\cite{montanha2023revisiting} following Queiroz et al. and Andreotti et al. methodologies using a realistic VH created in the state-of-art MetaHuman creator\footnote{https://www.unrealengine.com/en-US/metahuman}. In this study, Montanha and colleagues analyzed the perception of macro, micro, and macro followed by micro expressions in VHs and compared the results with the based studies. In these studies, the facial animations were made in an animation tool using the active AUs of each emotion, as presented in Ekman and Friesen's work. In 2011, Zielke et al.~\cite{zielke2011creating} discussed technical challenges and solutions in creating facial and micro expressions. At that time, the authors concluded that processes of driven facial animations of micro expressions required post-processing carried out by artists. Perhaps because technologies in 2011 did not have as much power to transfer facial features from a real person in a video to a VH animation in real-time. With the results of the presented literature on VHs micro and macro expressions, we raised two questions:
\textit{i)} Do the state-of-art techniques accurately transfer the macro and micro expressions from a real human to a VH? And, \textit{ii)} do people perceive similar emotions transferred from a real human and the emotion made by artists using an animation tool?

To answer the questions, this work extends Montanha et al.~\cite{montanha2023revisiting}, studying how people perceive realistic VH facial expressions transferred from a real human. Inspired by the same methodology of Queiroz et al.~\cite{queiroz2014investigating} and Andreotti et al.~\cite{andreotti2021perception}, we elaborated three hypotheses to be answered in the present work:

\begin{itemize}
    \item $H0_1$ - People similarly recognize and evaluate intensities of macro and micro emotional expressions, whether created using facial modeling tools or facial-driven animations.
    \item $H0_2$ - People feel similarly comfortable with macro and micro emotion expressions in driven facial animations. 
    \item $H0_3$ - People can identify macro and micro emotion expressions created through driven facial animation performed by people in real videos.
\end{itemize}

To test these hypotheses, we created a realistic VH in MetaHuman Creator, as in Montanha et al.~\cite{montanha2023revisiting}, and recorded the macro and micro expressions using the Live Link Face technology and two datasets: the DISFA+~\cite{mavadati2016extended} for macro expressions, and the SAMM~\cite{davison2016samm} for micro expressions. Our main goal in this work is to observe whether the macro and micro facial expressions of a VH animated from a real human being motion are perceived similarly to previous studies when the facial expressions were modeled in animation. Therefore, to our knowledge, this is the first time a perceptual study has qualitatively and statistically compared modeled versus driven animated micro and macro facial expressions.

\section{Related Work}
\label{sec:related_work}

As the center of human communication, faces enable humans to recognize even the slightest behavioral movements. Several studies have attempted to comprehend the expression of emotions, which is a natural human behaviour~\cite {bassili1978facial,Ekman2013AnAF,EkmanFACS,friesen1983emfacs}. Regarding micro expressions, the pioneering work of Gottschalk et al.~\cite{gottschalk1966micromomentary} aimed to understand the nonverbal communication between therapists and patients. Their study observed that the expression on the patient’s face would occasionally undergo drastic changes within three to five frames, equivalent to a period of 1/8 to 1/5 of a second, transitioning from a smile to a frown and back to a smile. Later, Ekman formally coined the term ``micro expressions".

Many studies in Computer Graphics (CG) emerge to understand better the facial animation~\cite{carrigan2020investigating,lewis2014practice,queiroz2014investigating}. From the blendshapes techniques~\cite{lewis2014practice} to facial tracking~\cite{barros2019face}, the goal is to provide realistic facial animations for VHs. However, some facial tracking methods, even though they use tracking tools such as the Kinect system of Microsoft, show limitations due to hardware and the small number of blend shapes in some models~\cite{weise2011realtime}. There is a limitation in the accuracy of transferring real expressions to computer-generated (CG) faces due to the smoothing intensities used in the process. According to a study by Peres et al.~\cite{peres2023can}, when reconstructing 3D faces expressing happiness and comparing the use of AU6 and AU12 (which are used to express happiness) in the CG character with real human expressions using OpenFace~\cite{amos2016openface}, the results were not accurate. Additionally, McDonnell et al.~\cite{mcdonnell2021model} examined how gender and race can affect the perception of AUs and blendshapes.

Some researchers also studied how people perceived emotions in VHs~\cite{
bailey2017gender,ennis2013emotion,hyde2014assessing,hyde2016evaluating,katsyri2015review,ochs201518,queiroz2014investigating,zibrek2013evaluating,montanha2023revisiting}. Ennis et al.~\cite{ennis2013emotion} conducted a perceptual study using synchronized full-body and facial driven animation data. The results indicate that individuals can recognize emotions from either body or facial motion alone. The authors evaluated four macro expressions: Anger, Fear, Happiness, and Sadness. The results also indicated that Happiness had the highest percentage of correct answers, with a percentage of correct answers above 70\% of the participants, while Fear had the lowest percentage (below 40\%). Hyde et al. study~\cite{hyde2014assessing} explored how modifying auditory and facial expressiveness levels could affect accuracy in emotion recognition, perceived emotional intensity, and naturalness ratings. In the experiment, participants assessed animations featuring a character whose facial movements aligned with a monitored actress's. The results indicated that increased auditory expressiveness positively impacted both emotion recognition accuracy and emotional intensity ratings.

This present work builds upon the research conducted by Montanha et al.~\cite{montanha2023revisiting}, which was inspired by the studies of Queiroz et al.~\cite{queiroz2014investigating} and Andreotti et al.~\cite{andreotti2021perception}. The Queiroz et al.~\cite{queiroz2014investigating} study focused on three psychological experiments that investigated the perception of macro and micro expressions. One of these experiments, proposed by Bornemann et al.~\cite{bornemann2012can}, involved showing micro expressions lasting between 10 and 20 milliseconds to ensure that individuals did not consciously perceive them. The other two experiments, the Brief Affect Recognition Test (BART) and the Micro Expression Training Tool (METT), were presented in Shen et al.'s work~\cite{shen2012effects}, following the studies of Ekman. In the BART scenario, participants were shown the six universal expressions in succession at a fixed point, while in the METT paradigm, the universal expressions were displayed amongst two sequences of neutral faces. The research aimed to determine the maximum duration required for detecting micro expressions. The study found that participants' response accuracy stabilized at 160 milliseconds.
Inspired by Queiroz et al.~\cite{queiroz2014investigating}, Andreotti et al.~\cite{andreotti2021perception} used these three psychological studies to study micro and macro expressions at a different level of realism. While Queiroz and collaborators used a realistic VH, Andreotti used a cartoon VH.

Other studies involved micro expressions and VHs. For example, in the work of Tastemirova et al.~\cite{tastemirova2022microexpressions}, the authors used the SAMM dataset as a visual base to create micro expression animations in VHs. Unlike our work, we also used two datasets (both SAMM and DISFA+ datasets) to create driven facial animations. The authors evaluated how micro expressions of happiness and anger in realistic VHs impacted perceptions of affection, sincerity, trustworthiness, eeriness, humanness, and attractiveness. In some results, the authors concluded that the intensity of the micro expression influenced perception. Furthermore, the authors qualitatively evaluated micro expression transfers from facial features extracted using two AI-based techniques (GANimation~\cite{Pumarola_ijcv2019} and NVIDIA fs‑vid2vid~\cite{wang2019fewshotvid2vid}). The authors concluded that modeled animation methods were better at representing micro expressions. However, unlike our work, the authors did not use real-time driven facial animations and also did not use statistical analysis in that evaluation. In the work of Hout et al.~\cite{hou2022micro}, the authors examined whether people could recognize driven animations of micro expressions in cartoon VHs, whether the intensity of micro expressions would affect emotion recognition, and whether there were differences in recognition and perception based on the visual style of the VH (realistic vs. cartoon). The results showed that the intensity of the micro expressions was important to emotion recognition, and this was more noticeable in the realistic VH than in the cartoon VH.

However, as we have seen in all studies involving micro expressions in VH, and to our knowledge, no previous study has compared micro expressions in modeled animated versus driven-animated VHs. Furthermore, no study has evaluated micro expressions versus macro expressions in driven-animated VHs. 
Table~\ref{tab:relatedWorkComparison} presents the related studies in this section.

\begin{table}[!htp]
    \centering
    \footnotesize
    \begin{tabular}{|c|c|c|c|}
    \hline
    \textbf{Work} & \textbf{VH Realism} & \textbf{Micro and Macro} & \textbf{Animation Type}   \\ \hline
    Montanha et al.~\cite{montanha2023revisiting} & Realistic & Both & Modeled Animated             \\  \hline
    Queiroz et al.~\cite{queiroz2014investigating} 
                           & Realistic & Both & Modeled Animated             \\ \hline 
                           Andreotti et al.~\cite{andreotti2021perception} & Cartoon &  Both   & Modeled Animated             \\ \hline Tastemirova et al.~\cite{tastemirova2022microexpressions}
                           & Realistic &                Micro   & Modeled Animated             \\ \hline Hou et al.~\cite{hou2022micro}
                           & Cartoon &   Micro                   & Modeled Animated  \\ \hline
    \end{tabular}
    \caption{The table presents related studies on macro and micro expressions. Furthermore, the table also presents the type of realism of the VHs and the type of animation used in each study.}
    \label{tab:relatedWorkComparison}
\end{table}

\section{Methodology}
\label{sec:methodology}

The methodology proposed in this work aims to investigate the following hypotheses: $H0_1$ (People recognize emotions similarly, whether the VHs are created using an animation tool or performance-driven animation), $H0_2$ (People feel comfortable while watching VHs facial motions created using performance-driven animation), and $H0_3$ (People correctly identify the emotions expressed by the virtual humans created using performance-driven animation). Firstly, we discuss the stimuli and present how we proceed with the emotion transfer from real to virtual humans. Then, we describe the questionnaire used to collect answers from the subjects.

\subsection{Stimuli Creation}
\label{sec:stimuli}

Unlike previous studies~\cite{queiroz2014investigating,andreotti2021perception,montanha2023revisiting} that evaluated the impact of facial micro and macro expressions in human perception when modeled by professionals such as artists, animators, and designers, we aim to investigate the transfer of real faces to virtual ones through performance-driven facial animations. To achieve this, we used Live Link Face facial tracking technology\footnote{https://dev.epicgames.com/community/learning/tutorials/lEYe/unreal-engine-facial-capture-with-live-link~\label{foot:live-link}} to transfer facial expressions from a real human to the VH. For this purpose, we required videos where actors express micro and macro expressions with annotated emotions. Therefore, we selected two datasets, DISFA+~\cite{mavadati2016extended} for macro expressions and SAMM~\cite{davison2016samm} for micro expressions. These datasets contain videos of real humans expressing the six Ekman's expressions~\cite{Ekman2013AnAF} and emotions attributed and validated by the authors.

Initially, we divided the stimuli into two parts. In the first part, we utilized videos of one actor - expressing all six emotions through micro and macro facial expressions. The 12 videos containing the actor were transferred to a VH, resulting in 12 short sequences in CG. The first part of the analysis evaluates how subjects answer for recognition, emotion intensity, and perceived comfort, forming the basis for Hypotheses $H0_1$ and $H0_2$. So, this analysis employed a design involving two factors: micro and macro expressions and six emotions (happiness, sadness, fear, anger, disgust, and surprise), each evaluated across three criteria: recognition, intensity, and comfort. This first part of the experiment results in 12 videos with VHs. 
6 videos from DISFA+\footnote{Macro Expressions First Part - Video file names (please, see DISFA+~\cite{mavadati2016extended} dataset for more details): SN007 Z Happiness Text Trail No 1; SN007 Z Sadness Text Trail No 2; SN007 Z Surprise Text Trail No 1; SN007 Z Fear Text Trail No 1; SN007 Z Disgust Text Trail No 1; SN007 Z Anger Text Trail No 2} regarding macro expressions and 6 videos from SAMM regarding micro expressions\footnote{Micro Expressions First Part - Video file names (please, see SAMM~\cite{davison2016samm} dataset for more details): Happiness 011 6 6; Sadness 033 1 4; Surprise 014 2 3; Fear 035 6 5; Disgust 018 3 1; Anger 006 5 11}.

For the second part of our study (related to $H0_3$), we narrowed our focus to two emotions: Happiness and Anger. These emotions were selected based on their high recognition rates, as indicated by a study conducted by Savage et al. \cite{savage2013search}. Indeed, to ensure clarity in our analysis and avoid potential interference from other variables, we deliberately excluded additional emotions from this portion of the study. As in the first part, we studied the macro and micro facial expressions. For macro expressions, we utilized two videos\footnote{Macro Expressions Second Part - Video file names (please, see DISFA+ dataset for more details): SN007 Z Anger Text Trail No 2; SN00 Z Happiness Text Trail No 1} featuring two real actors for each analyzed emotion (4 videos with real actors) alongside the corresponding videos transferred to VHs, resulting in 4 videos of VHs.
However, in the questionnaire, we presented only one video featuring a real actor alongside the two computer-generated (CG) videos for each emotion, resulting in six videos. We aimed to assess user perception and determine the subjects' ability to accurately associate real and virtual expressions. 
For micro expressions, we utilized a single actor and transferred his/her micro expressions of the two emotions\footnote{Micro Expressions Second Part - Video file names (please, see SAMM dataset for more details): Happiness 011 6 5; Anger 006 5 11} and the neutral expression to VHs, resulting in 3 videos. Once more, we aimed to ascertain whether the transfer was sufficiently effective for subjects to perceive even the most subtle facial expressions. More details about the questionnaire are presented in Section~\ref{sec:questionnaire}.

To guarantee the authenticity of emotions depicted in our dataset, we meticulously chose actors for their emotional depth, in our empirical opinion. Employing visual assessments, we excluded individuals whose emotional expressions lacked the requisite intensity. Consequently, only videos featuring the most emotionally intense actor were incorporated into our dataset. Admittedly, this selection process presents a limitation in our study; however, given the substantial time required for subjects to watch and analyze each video, we faced constraints on including more than one actor for inclusion in the survey.
\begin{table}[!htp]
    \centering
    \footnotesize
    \begin{tabular}{|c|c|c|c|c|}
    \hline
    \textbf{Macro / Micro} & \textbf{Emotion}& \textbf{Real} & \textbf{CG}  & \textbf{p-value} \\ \hline
    \multirow{6}{*}{Macro} & Happiness & $0.41$ & $0.20$ & $<.001$             \\  \cline{2-5} 
                           & Sadness                     & $0.27$ & $0.09$ & $<.001$             \\ \cline{2-5} 
                           & Anger  & $0.32$ &  $0.10$   & $<.001$             \\ \cline{2-5} 
                           & Surprise  & $0.17$ &                $0.10$   & $<.001$             \\ \cline{2-5} 
                           & Fear  & $0.29$ &   $0.10$                   & $<.001$             \\ \cline{2-5} 
                           & Disgust  & $0.46$ & $0.16$                  & $<.001$             \\ \hline
    \multirow{6}{*}{Micro} & Happiness                   & $0.31$ & $0.23$ & $<.001$             \\ \cline{2-5} 
                           & Sadness                    & $0.19$ & $0.17$ & $.06$*              \\ \cline{2-5} 
                           & Anger                      & $0.16$ & $0.12$ & $<.001$             \\ \cline{2-5} 
                           & Surprise                   & $0.12$ & $0.10$ & $.001$              \\ \cline{2-5} 
                           & Fear                       & $0.24$ & $0.18$ & $<.001$              \\ \cline{2-5} 
                           & Disgust                    & $0.42$ & $0.34$ & $<.001$              \\ \hline
    \end{tabular}
    \caption{Table presents the average intensity of the AUs generated by the OpenFace software for the real videos (DISFA+ dataset~\cite{mavadati2016extended} on macro expressions, and SAMM dataset~\cite{davison2016samm} on micro expressions) and for the CG (driven animations) videos. Furthermore, the table presents the p-values of the \textit{T}-test for comparisons of average intensities between real video and CG video. The * highlights the only non-significant \textit{p}-value, considering a significance level of 5\%.}
    \label{tab:videos}
\end{table}

\subsubsection{Evaluating the Facial Expression Transfer}

This analysis aimed to measure statistically the similarity between real and computer-generated faces based on Action Unities (AUs) activation and intensities. This analysis helps to determine if, at least quantitatively,  the videos involving VHs successfully represented expressions (micro and macro) in the same way as real videos. We used the OpenFace software~\cite{baltrusaitis2018openface, baltruvsaitis2015cross}  to obtain intensity values for the facial AUs in the selected videos.
OpenFace is a Computer Vision software focused on facial behavior analysis in videos. This tool provides objective values for each frame related to facial landmarks, such as AU activations, intensities, and other features. In our work, we ran OpenFace on all 
the videos of the datasets (the videos containing real people) and also on the CG videos generated from the real ones. From the features generated by OpenFace, we used only the intensity values of AUs 2, 4, 5, 12, 15, and 26, which according to studies by Mcdonnell et al.~\cite{mcdonnell2021model} and Carrigan et al.~\cite{carrigan2020investigating}, are the important AUs to represent emotions. Still according to these studies, AU 38 was also important for emotions, however, OpenFace does not generate values for this AU. So, using independent \textit{T}-test (significance level of 5\%), we performed pairwise comparisons of intensities between the CG videos versus the dataset videos. The average intensities and \textit{p}-values for comparisons between them can be seen in Table~\ref{tab:videos}.

Regarding the results of the analyses, as we can see in Table~\ref{tab:videos}, among all the comparisons between CG video versus dataset video, only the comparison involving the micro expression of sadness did not have a significant result (Stats=-1.83 and \textit{p}-value=$.06$). Furthermore, all the average intensities of the videos with actors in the dataset were greater than those of the transferred CG videos. So, in general, the micro and macro expressions transfers from real to CG were significantly different. However, we still decided to continue using CG videos as a stimulus for our research because we wanted to know how human perception would react as a function of those differences.

\subsubsection{Reconstruction of the Virtual Human}

\begin{figure*}[!htp]
    \centering
    \subfigure[Anger]{
        \includegraphics[width=0.20\textwidth]{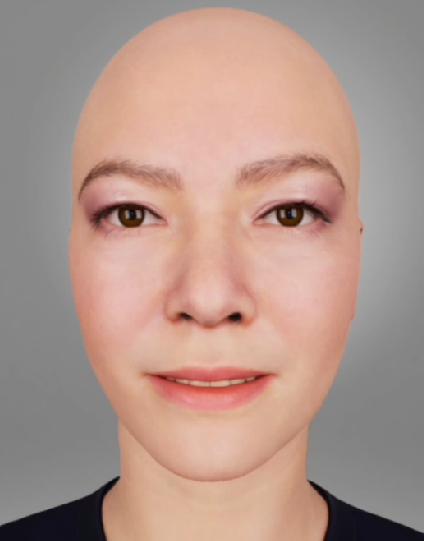}
        \label{fig:anger-macro}
    }
    \subfigure[Disgust]{
        \includegraphics[width=0.20\textwidth]{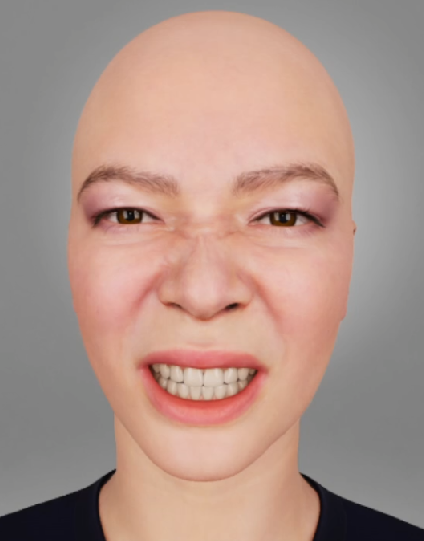}
        \label{fig:disgust-macro}
    }
    \subfigure[Fear]{
        \includegraphics[width=0.20\textwidth]{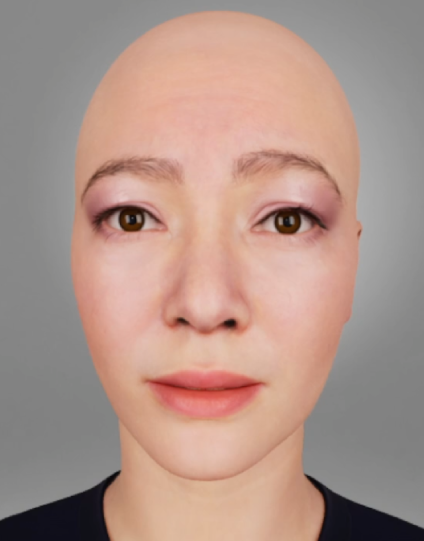}
        \label{fig:fear-macro}
    } \\
    \subfigure[Happiness]{
        \includegraphics[width=0.20\textwidth]{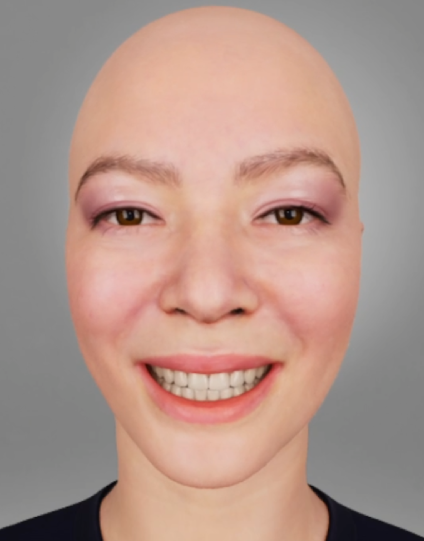}
        \label{fig:happiness-macro}
    }
    \subfigure[Sadness]{
        \includegraphics[width=0.20\textwidth]{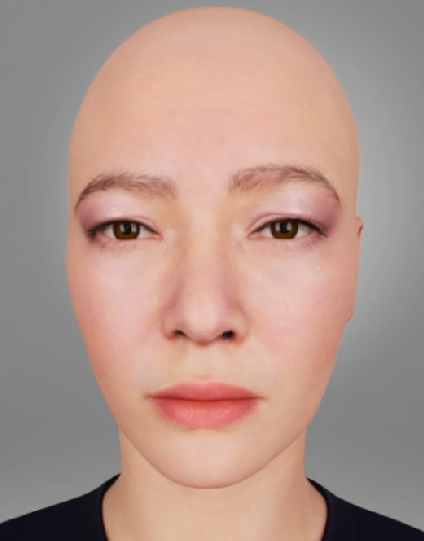}
        \label{fig:sadness-macro}
    }
    \subfigure[Surprise]{
        \includegraphics[width=0.20\textwidth]{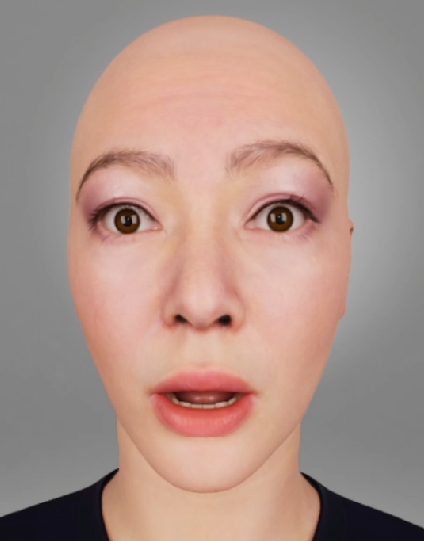}
        \label{fig:surprise-macro}
    }
    \caption{All macro expressions represented by the realistic character used in our work.}
    \label{fig:macro-expressions}
\end{figure*}

\begin{figure*}[!htp]
    \centering
    \subfigure[Anger]{
        \includegraphics[width=0.20\textwidth]{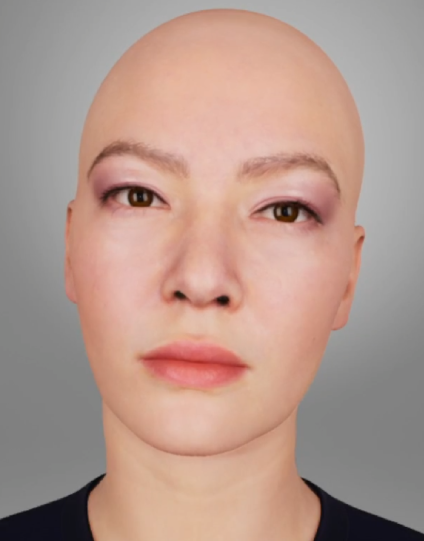}
        \label{fig:anger-micro}
    }
    \subfigure[Disgust]{
        \includegraphics[width=0.20\textwidth]{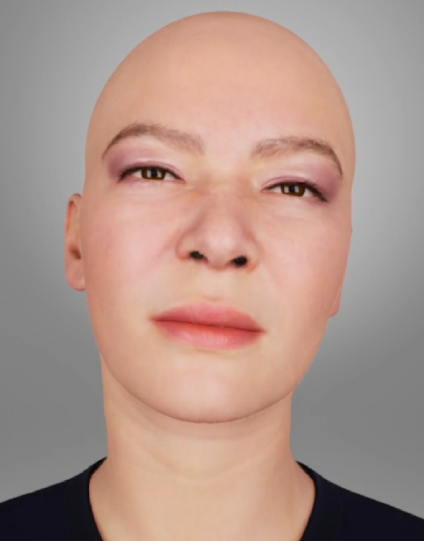}
        \label{fig:disgust-micro}
    }
    \subfigure[Fear]{
        \includegraphics[width=0.20\textwidth]{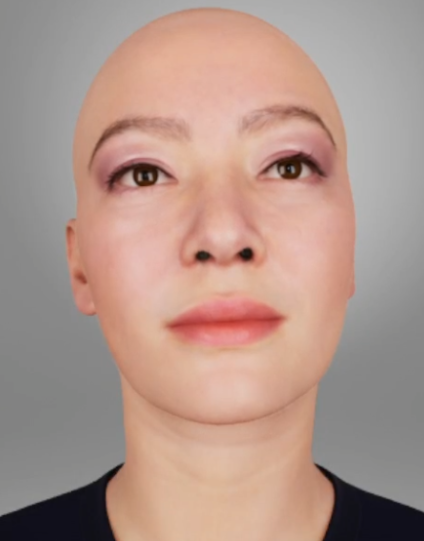}
        \label{fig:fear-micro}
    } \\
    \subfigure[Happiness]{
        \includegraphics[width=0.20\textwidth]{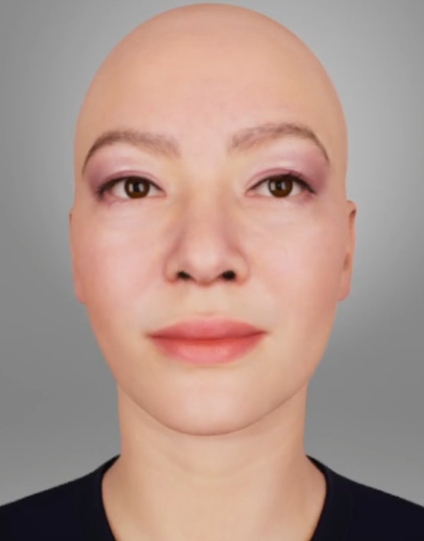}
        \label{fig:happiness-micro}
    }
    \subfigure[Sadness]{
        \includegraphics[width=0.20\textwidth]{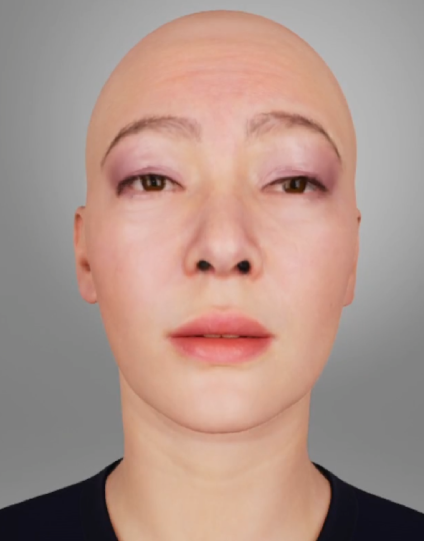}
        \label{fig:sadness-micro}
    }
    \subfigure[Surprise]{
        \includegraphics[width=0.20\textwidth]{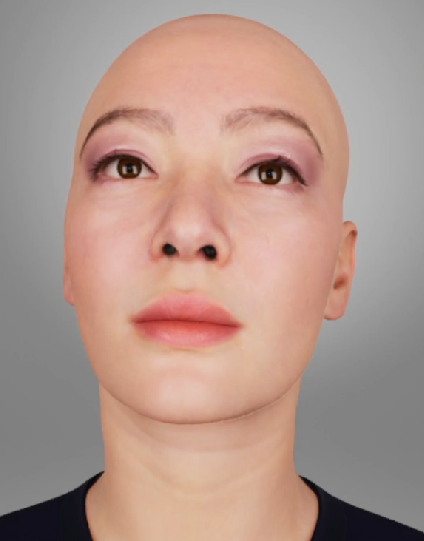}
        \label{fig:surprise-micro}
    }
    \caption{All micro expressions represented by the realistic character used in our work.}
    \label{fig:micro-expressions}
\end{figure*}

With respect to the VH reconstruction and to compare with the literature, we selected the same realistic female VH used in Montanha et al.'s previous work~\cite{montanha2023revisiting}. We used the MetaHuman Creator\footnote{https://www.unrealengine.com/en-US/metahuman} to create and animate high-fidelity VHs in Unreal Engine\footnote{https://www.unrealengine.com/ \label{foot:unreal}} $5.3$. The MetaHuman Creator offers a range of VHs to choose from, and to avoid any hair bias, we selected a hairless VH model, as done in the two baseline studies. Finally, we utilized the Quixel Bridge plug-in to export the VH to Unreal Engine for animation and facial expression creation. With the videos from DISFA+ and SAMM datasets, we used the Live Link Face\footref{foot:live-link} in an iPhone 13\footnote{https://www.apple.com/iphone-13/specs/} to capture and transfer the real actor's facial movements and video assets to VH within Unreal Engine from a computer monitor. This process allowed us to transfer the expressions made by actors 
to the VH in real-time, allowing us to record the animations from it. We carefully positioned the smartphone on a tripod in front of the computer monitor so that all micro and macro expression animations had the same recording process. For standardizing the video size of stimuli created, the VH began with a neutral expression, followed by micro or macro expressions, and returned to a neutral expression. Each video stimulus lasted between 8 and 10 seconds for macro expressions and 2 and 4 seconds for micro expressions. In addition, we used a white object for the video background that could cover the entire field of view behind the VH with standard lighting. Also, to capture all expressions, all videos were focused only on VH's face. The VH used in our work applying the recorded macro and micro expressions are illustrated in Figures~\ref{fig:macro-expressions} and \ref{fig:micro-expressions}, respectively. 

\subsection{Questionnaire}
\label{sec:questionnaire}

We conducted an online survey using the Qualtrics survey tool~\footnote{https://www.qualtrics.com/}. Please note that we have omitted information about the project due to blind submission. The survey was divided into four parts: consent form, demographic questions, experiment part 1, and experiment part 2. The consent form explained the purpose of the research to the participant, asks some demographic questions about personal information (gender, education, age), and familiarity with CG.

\begin{figure*}[!htp]
    \centering
    \includegraphics[width=0.6\textwidth]{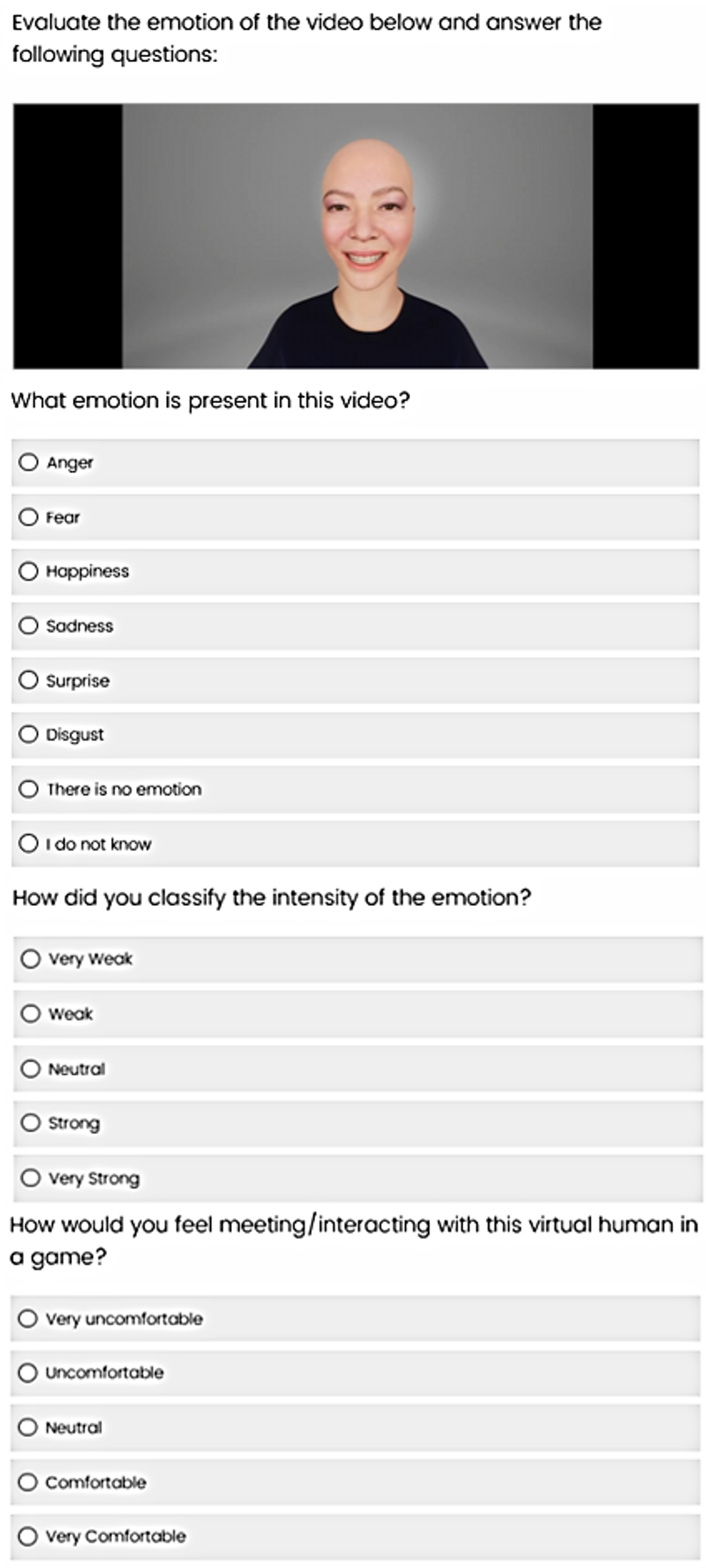}
    \caption{Figure of the first part of the survey. Firstly, it is presented a CG video, presenting one of the six basic emotions, presented in Figures~\ref{fig:macro-expressions} and \ref{fig:micro-expressions}, followed by the three questions outlined in Table~\ref{tab:questions-p1}.}
    \label{fig:survey-part-1}
\end{figure*}

The experiment part 1 was presented after the demographic questions, where the participant was invited to identify the emotions expressed by the VH. To do that, this part presented a sequence of the CG videos
. The first videos of this part are the six basic emotions (Anger, Disgust, Fear, Happiness, Sadness, and Surprise) expressed through macro expressions in a random order for each participant. After the macro expressions, a series of six videos of micro expressions, one of each emotion, was presented to the participant, randomly ordered as it was done for the macro expressions. After each video, both for macro and micro expressions, the three questions of Table~\ref{tab:questions-p1} were asked to the participant, as shown in Figure~\ref{fig:survey-part-1}. The question ``What emotion is present in this video?" was the same question used in the previous studies~\cite{andreotti2021perception,montanha2023revisiting}, and it was used to answer the hypothesis $H0_1$ in comparisons between the present study and the previous studies. Whereas the questions ``How did you classify the intensity of the emotion?" and ``How would you feel meeting/interacting with this virtual human in a game?" were used, respectively, to answer $H0_1$ and $H0_2$, in the present work.

\begin{figure*}[!htp]
    \centering
    \includegraphics[width=0.60\textwidth]{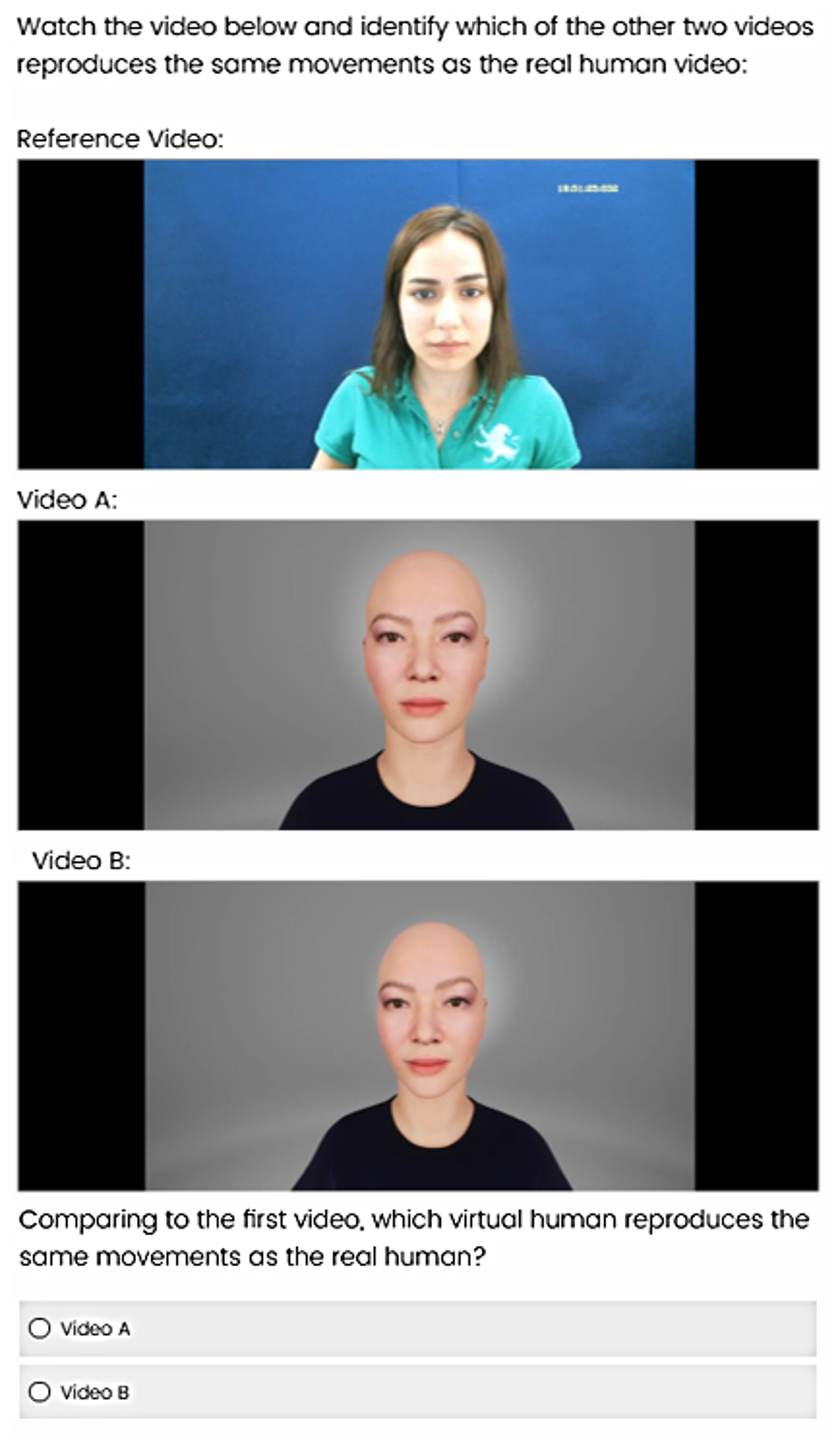}
    \caption{Figure of the second part of the survey. It starts with three videos, followed by a question. One of the three videos features a real human, while the other two are VHs. For macro expressions, both videos express the same emotion but are transferred from different actors, one of them the same as the reference video. For micro expressions, one video is the animation transferred from the reference video, while the other is a neutral face. The reference video used in this example is from DISFA+ dataset~\cite{mavadati2016extended}.}
    \label{fig:survey-part-2}
\end{figure*}

For the experiment part 2, four cases were presented to the participants. Two cases related to macro expressions (for anger and happiness emotions) and two cases related to micro expressions (also for anger and happiness emotions). In each case, three videos were shown on the screen to the participants, and 
accompanied by a question. One of the three videos contained a real human, while the other two contained a VH. For macro expressions, both videos containing VHs expressed the same emotion, but in one video, the VH was animated with the same movements as the video presented with the real person, and in the other video, the VH was animated with an actor performance video not shown to the participants. For micro expressions, one video had the VH animated with the reference actor, while in the other video, the VH presented a neutral face. The only question asked in this part, as presented in Table~\ref{tab:questions-p2}, was ``Comparing to the first video, which virtual human reproduces the same movements as the real human?" and aims to answer the hypothesis $H0_3$. Figure~\ref{fig:survey-part-2} shows an example from the survey.

\begin{table}[!htb]
    \centering
    \begin{tabular}{|c|c|}
    \hline
    \textbf{Question}                                                                                                         & \textbf{Possible Answers}                                                                                                           \\ \hline
    \begin{tabular}[c]{@{}c@{}}Q1: What emotion is \\ present in this video?\end{tabular}                                     & \begin{tabular}[c]{@{}c@{}}Anger; Fear; Happiness; \\ Sadness; Surprise;Disgust; \\ There is no emotion; I do not know\end{tabular} \\ \hline
    \begin{tabular}[c]{@{}c@{}}Q2: How did you classify \\ the intensity of the emotion?\end{tabular}                         & \begin{tabular}[c]{@{}c@{}}5-Likert from Very Weak\\ to Very Strong\end{tabular}                                                    \\ \hline
    \begin{tabular}[c]{@{}c@{}}Q3: How would you feel\\ meeting/interacting with\\ this virtual human in a game?\end{tabular} & \begin{tabular}[c]{@{}c@{}}5-Likert from very uncomfortable\\ to very comfortable\end{tabular}                                      \\ \hline
    \end{tabular}
    \caption{The table presents the questions applied in the first part of the questionnaire to the participants. The ``Question" column provides the questions used in the questionnaire, while the ``Possible Answers" column lists the options available for the participants.}
    \label{tab:questions-p1}
\end{table}

\begin{table}[!htb]
    \centering
    \begin{tabular}{|c|c|}
    \hline
    \textbf{Question}                                                                                                                                   & \textbf{Possible Answers}                                                                 \\ \hline
    \begin{tabular}[c]{@{}c@{}}Q1: Comparing to the first video,\\  which virtual human reproduces\\ the same movements as the real human?\end{tabular} & Video A; Video B                                                                          \\ \hline
    
    \end{tabular}
    \caption{The table presents the questions applied to the participants in the second part of the questionnaire. The ``Question" column provides the questions used in the questionnaire, while the ``Possible Answers" column lists the options available for the participants.}
    \label{tab:questions-p2}
\end{table}

\section{Results}
\label{sec:results}
This section aims to present the results of our work. The questionnaire was distributed on social media, and 73 participants volunteered. Of these participants, 46 (63.0\%) were females, while 26 (35.6\%) were males, and one preferred not to answer. About age, 45.2\% were between 18 and 20 years, followed by the group between 21 and 30 years (35.6\%). Most of them complete their secondary education (67.1\%). Focusing on CG experience, most people declared a low (31.5\%) or very low (28.7\%) experience, while just 13.6\% declared a high or a very high level of experience. The others 26.0\% declared a normal level of previous experience.

Therefore, this section is structured as follows: \textit{i)} Section~\ref{sec:comparisonStudiesResults} presents the results referring to question $Q1$ of Table~\ref{tab:questions-p1} to test the $H0_1$ in the comparison between our work and studies in the literature (\cite{montanha2023revisiting,queiroz2014investigating,andreotti2021perception,hou2022micro}); 
\textit{ii)} Section~\ref{sec:intensityResults} presents the findings referring to $Q2$ (Table~\ref{tab:questions-p1}) with the aim of testing $H0_1$ concerning the perception of the intensity of driven animated micro and macro expressions; \textit{iii)} Section~\ref{sec:comfortResults} presents the results of $Q3$ (Table~\ref{tab:questions-p1}) with the aim to test $H0_2$ regarding the perception of comfort; and \textit{iv)} Section~\ref{sec:comparisonVideosResults} provides the results related to $Q1$ of Table~\ref{tab:questions-p2} to test $H0_3$ in identifying which emotion presented by VH in the CG video was the emotion presented in the real video.

\begin{table*}[!htb]
    \scriptsize
    \centering
    \begin{adjustbox}{max width=0.98\linewidth}
    \begin{tabular}{|c|c|c|c|c|c|c|c|c|}
    \hline
    \textbf{\begin{tabular}[c]{@{}c@{}}Macro  \\ \end{tabular}} & \textbf{Happy} & \textbf{Sad} & \textbf{Anger}  & \textbf{Surprise} & \textbf{Fear}    & \textbf{Disgust} & \textbf{\begin{tabular}[c]{@{}c@{}}No\\ Emotion\end{tabular}} & \textbf{\begin{tabular}[c]{@{}c@{}}Don't \\ Know\end{tabular}} \\ \hline
    \textbf{Happy}                                                 & \textbf{95.89\%}   & 1.36\%           & 0.00\%          & 0.00\%            & 0.00\%           & 1.36\%           & 0.00\%                                                        & 1.36\%                                                           \\ \hline
    \textbf{Sad474}                                                   & 0.00\%             & \textbf{27.39\%} & 4.10\%          & 19.17\%           & 5.47\%           & 20.54\%          & 5.47\%                                                        & 17.90\%                                                          \\ \hline
    \textbf{Anger}                                                     & 41.09\%            & 0.00\%           & \textbf{1.36\%} & 6.84\%            & 6.84\%           & 8.21\%           & 13.69\%                                                       & 21.91\%                                                          \\ \hline
    \textbf{Surprise}                                                  & 0.00\%             & 0.00\%           & 1.36\%          & \textbf{89.04\%}  & 8.21\%           & 0.00\%           & 1.36\%                                                        & 0.00\%                                                           \\ \hline
    \textbf{Fear}                                                      & 6.84\%             & 21.91\%          & 0.00\%          & 17.80\%           & \textbf{20.54\%} & 13.69\%          & 2.73\%                                                        & 16.43\%                                                          \\ \hline
    \textbf{Disgust}                                                   & 1.36\%             & 0.00\%           & 67.12\%         & 0.00\%            & 1.36\%           & \textbf{23.28\%} & 2.73\%                                                        & 4.10\%                                                           \\ \hline
    \end{tabular}
    \end{adjustbox}
    \caption{Confusion matrix displaying the percentages of all response options (Happiness, Sadness, Surprise, Fear, Anger, Disgust) for facial macro 
    expressions. Participants' responses are presented along the extensions of the columns following the top row. These responses correspond to Q1 of Table~\ref{tab:questions-p1} related to macro expressions, as presented in Section~\ref{sec:questionnaire}.}
    \label{tab:macro-matrix}
\end{table*}

\begin{table}[!htb]
    \scriptsize
    \centering
    \begin{adjustbox}{max width=0.98\linewidth}
    \begin{tabular}{|c|c|c|c|c|c|c|c|c|}
    \hline
    \textbf{\begin{tabular}[c]{@{}c@{}}Micro\\ \end{tabular}} & \textbf{Happy} & \textbf{Sad} & \textbf{Anger}   & \textbf{Surprise} & \textbf{Fear}   & \textbf{Disgust} & \textbf{\begin{tabular}[c]{@{}c@{}}No\\ Emotion\end{tabular}} & \textbf{\begin{tabular}[c]{@{}c@{}}Don't \\ Know\end{tabular}} \\ \hline
    \textbf{Happy}                                                & \textbf{50.68\%}   & 1.36\%           & 0.00\%           & 1.36\%            & 0.00\%          & 4.10\%           & 35.61\%                                                       & 6.84\%                                                           \\ \hline
    \textbf{Sad}                                                  & 0.00\%             & \textbf{8.21\%}  & 2.73\%           & 5.47\%            & 5.47\%          & 6.84\%           & 41.09\%                                                       & 30.13\%                                                          \\ \hline
    \textbf{Anger}                                                    & 0.00\%             & 1.36\%           & \textbf{41.09\%} & 5.47\%            & 2.73\%          & 28.76\%          & 12.32\%                                                       & 8.21\%                                                           \\ \hline
    \textbf{Surprise}                                                 & 2.73\%             & 8.21\%           & 0.00\%           & \textbf{17.80\%}  & 2.73\%          & 0.00\%           & 58.90\%                                                       & 9.58\%                                                           \\ \hline
    \textbf{Fear}                                                     & 21.91\%            & 5.47\%           & 0.00\%           & 2.73\%            & \textbf{0.00\%} & 0.00\%           & 63.01\%                                                       & 6.84\%                                                           \\ \hline
    \textbf{Disgust}                                                  & 1.36\%             & 0.00\%           & 28.76\%          & 4.10\%            & 0.00\%          & \textbf{50.68\%} & 6.84\%                                                        & 8.21\%                                                           \\ \hline
    \end{tabular}
    \end{adjustbox}
    \caption{Confusion matrix displaying the percentages of all response options (Happiness, Sadness, Surprise, Fear, Anger, Disgust) for facial micro expressions. Participants' responses are presented along the extensions of the columns following the top row. These responses correspond to Q1 of Table~\ref{tab:questions-p1} related to micro expressions, as presented in Section~\ref{sec:questionnaire}.}
    \label{tab:micro-matrix}
\end{table}

\subsection{Comparisons on Emotion Recognition between Driven Animation versus Modeled Animation - $H0_1$}
\label{sec:comparisonStudiesResults}

To answer $H0_1$, we needed to compare our $Q1$ results (that is, performance-driven facial animations) versus the results of the studies by Andreotti et al.~\cite{andreotti2021perception}, Queiroz et al.~\cite{queiroz2014investigating}, and Montanha et al.~\cite{montanha2023revisiting} (that is, modeled animation). 

Firstly, looking at the percentages of correct responses in our work about macro expressions, as we can see in the bold values of Table~\ref{tab:macro-matrix}, happiness and surprise had percentages of correct responses close to 90\%. All other four emotions had correct response percentages below 30\%. Some participants were confused in recognizing some of these four emotions, with 41\% of participants thinking that the macro expression of anger was happiness, and 67\% thought that the emotion of disgust was anger. Concerning micro expressions, as we can see in the bold values of Table~\ref{tab:micro-matrix}, half of the participants had correct answers in relation to the happiness and disgust micro expressions, and 40\% in relation to the anger micro expression. In each of the other emotions, less than 20\% of participants had correct answers. 40\% of the participants in each emotion responded that no emotion was presented in the video.

Comparing the work of Montanha et al. (presented in Table~\ref{tab:macroStudiesTab}) and our macro expression results, we can see that the average percentage of correct answers in the work of Montanha et al. was higher than the average percentage of our work. Looking at the percentages of correct answers of each emotion, our work had the highest percentage related to happiness and surprise and the lowest percentages related to other emotions.  So, \textbf{we rejected $H0_1$ once analyzing only the percentages of correct answers, we observed that the values are different in the comparisons on macro expressions between driven animations and modeled animations.} It is important to note that due to the missing specific values of emotions of other work in the literature, we could not do the statistical analysis.

\begin{table}[!htp]
    \scriptsize
    \centering
    \begin{adjustbox}{max width=0.98\linewidth, center}
    \begin{tabular}{|c|c|c|c|c|c|c|c|}
    \hline
    \textbf{Macro}                                                            & \textbf{Happy} & \textbf{Sad} & \textbf{Anger}   & \textbf{Surprise} & \textbf{Fear}    & \textbf{Disgust} & \textbf{Avg (\%)} \\ \hline
    \textbf{\begin{tabular}[c]{@{}c@{}}Our\\  - Realistic\end{tabular}}                & \textbf{95.89\%}            & {\ul27.39\%}           & {\ul1.36\%}    &  \textbf{89.04\%}    & {\ul20.54\%}   &   {\ul23.28\%}  & {\ul42.91\%}      \\ \hline
    \textbf{\begin{tabular}[c]{@{}c@{}}Montanha et al.~\cite{montanha2023revisiting} \\ - Realistic\end{tabular}}    & {\ul90.00\%}            & \textbf{80.00\%} & \textbf{83.33\%}          & {\ul83.33\%}           & \textbf{80.00\%}          & \textbf{80.00\%} & \textbf{82.77\%} \\ \hline
    \textbf{\begin{tabular}[c]{@{}c@{}}Andreotti et al.~\cite{andreotti2021perception} \\ - Unrealistic\end{tabular}} & -     & -     & -         & -           & -         & - & 80.00\%   \\ \hline
    \end{tabular}
    \end{adjustbox}
    \caption{Percentages of all emotions (Happiness, Sadness, Surprise, Fear, Anger, Disgust) of macro facial expressions in relation to the three studies (Our, Montanha et al.~\cite{montanha2023revisiting}, and Andreotti et al.~\cite{andreotti2021perception}). Queiroz et al. ~\cite{queiroz2014investigating} did not have only analysis with only macro expressions in stimuli. Bold values are the higher values, while the underlined values are the lower. }
    \label{tab:macroStudiesTab}
\end{table}

When we talk about micro expressions (Table~\ref{tab:microStudiesTab}), we found two significant results in the comparisons between our work and the studies by Queiroz et al. (\textit{U-statistic}=$2.0$ and \textit{p}=$.008$) and Montanha et al. (\textit{U-statistic}=$5.0$ and \textit{p}=$.04$), and no significant result in the comparison between our work and the work by Andreotti et al. (\textit{U-statistic}=$14.0$ and \textit{p}=$.58$). In the two cases where there were significant results, the people in the studies by Queiroz et al. and Montanha et al. had higher average percentages of correct answers than the people in our work. Looking only at the percentages presented in Table~\ref{tab:microStudiesTab}, our work did not have any of the highest percentages, and had the lowest percentages when they were micro expressions of anger, surprise, and fear. We can also compare the percentages of correct answers of our micro expressions results and in the work of Hou et al.~\cite{hou2022micro}, which had percentages related only to the micro expressions of happiness, sadness, fear and surprise in a cartoon VH (respectively, 84.73\%; 88.73\%; 60.73\%; 83.64\%, with an average of 79.45\%), and a realistic VH (respectively, (87.37\%; 82.94\%; 69.62\%; 77.13\%, with an average of 79.27\%). So, \textbf{about micro expressions, we reject $H0_1$ when comparing driven animations from realistic VHs versus modeled animations from realistic VHs. In this case, the recognition of micro expressions was more accurate when the VHs had modeled animations.}

\begin{table}[!htp]
    \scriptsize
    \centering
    \begin{adjustbox}{max width=0.98\linewidth, center}
    \begin{tabular}{|c|c|c|c|c|c|c|c|}
    \hline
    \textbf{Micro}                                                            & \textbf{Happy} & \textbf{Sad} & \textbf{Anger}   & \textbf{Surprise} & \textbf{Fear}    & \textbf{Disgust} & \textbf{Avg (\%)} \\ \hline
    \textbf{\begin{tabular}[c]{@{}c@{}}Our\\  - Realistic\end{tabular}}                & 50.68\%            & 8.21\%           & {\ul 41.09\%}    & {\ul 17.80\%}     & {\ul 0.00\%}     & 50.68\%  & {\ul 28.07\%}       \\ \hline
    \textbf{\begin{tabular}[c]{@{}c@{}}Montanha et al.~\cite{montanha2023revisiting} \\ - Realistic\end{tabular}}    & 83.33\%            & \textbf{80.00\%} & 50.00\%          & 90.00\%           & 36.67\%          & \textbf{86.67\%} & 71.11\% \\ \hline
    \textbf{\begin{tabular}[c]{@{}c@{}}Andreotti et al.~\cite{andreotti2021perception} \\ - Unrealistic\end{tabular}} & {\ul 43.20\%}      & {\ul 7.40\%}     & 61.70\%          & 69.10\%           & 16.00\%          & {\ul 30.90\%} & 38.05\%   \\ \hline
    \textbf{\begin{tabular}[c]{@{}c@{}}Queiroz et al.~\cite{queiroz2014investigating} \\ - Realistic\end{tabular}}     & \textbf{87.50\%}   & 75.60\%          & \textbf{92.26\%} & \textbf{92.86\%}  & \textbf{45.24\%} & 75.60\%  & \textbf{86.42\%}    \\ \hline
    \end{tabular}
    \end{adjustbox}
    \caption{Percentages of all emotions (Happiness, Sadness, Surprise, Fear, Anger, Disgust) of micro facial expressions in relation to the four studies (Our, Montanha et al.~\cite{montanha2023revisiting}, Andreotti et al.~\cite{andreotti2021perception}, and Queiroz et al~\cite{queiroz2014investigating}). Bold values are the higher values, while the underlined values are the lower.}
    \label{tab:microStudiesTab}
\end{table}

\subsection{Comparisons on Emotion Intensity Perception for Driven Animation - $H0_1$}
\label{sec:intensityResults}

Still about $H0_1$, about perception of emotions intensity in driven animations, the score created for question $Q2$ in Table~\ref{tab:questions-p1} was based on the 5 Likert scales, that is, from 1 to 5, with the answer options ``Very Weak" = 1 and ``Very Strong" = 5. In all analyses in this section, we use the Intensity Scale as the response variable and the Video Emotion and Type of Expression (micro or macro) as predictor variables. The significance level used was 5\%. For further details, some supplementary tables will be highlighted in \ref{sec:apendIntensity}.

\begin{table}[!htb]
    \centering
    \footnotesize
    \begin{tabular}{|cccccc|}
    \hline
    \multicolumn{6}{|c|}{\textbf{Emotion Intensity}}                                                                                                                                                                                                                                                       \\ \hline
    \multicolumn{1}{|c|}{\multirow{2}{*}{\textbf{Emotion}}} & \multicolumn{2}{c|}{\textbf{User’s Answer}}                                       & \multicolumn{1}{c|}{\multirow{2}{*}{\textbf{\begin{tabular}[c]{@{}c@{}}Macro / \\ Micro\end{tabular}}}} & \multicolumn{2}{c|}{\textbf{User’s Answer}}      \\ \cline{2-3} \cline{5-6} 
    \multicolumn{1}{|c|}{}                                  & \multicolumn{1}{c|}{\textbf{Avg}}       & \multicolumn{1}{c|}{\textbf{STD}}       & \multicolumn{1}{c|}{}                                                                                   & \multicolumn{1}{c|}{\textbf{Avg}} & \textbf{STD} \\ \hline
    \multicolumn{1}{|c|}{\multirow{2}{*}{Happiness}}        & \multicolumn{1}{c|}{\multirow{2}{*}{$2.90$}} & \multicolumn{1}{c|}{\multirow{2}{*}{$1.26$}} & \multicolumn{1}{c|}{Macro}                                                                              & \multicolumn{1}{c|}{\textbf{3.75}}            & $0.96$            \\ \cline{4-6} 
    \multicolumn{1}{|c|}{}                                  & \multicolumn{1}{c|}{}                   & \multicolumn{1}{c|}{}                   & \multicolumn{1}{c|}{Micro}                                                                              & \multicolumn{1}{c|}{$2.05$}            & $0.89$            \\ \hline
    \multicolumn{1}{|c|}{\multirow{2}{*}{Sadness}}          & \multicolumn{1}{c|}{\multirow{2}{*}{$2.26$}} & \multicolumn{1}{c|}{\multirow{2}{*}{$0.97$}} & \multicolumn{1}{c|}{Macro}                                                                              & \multicolumn{1}{c|}{$2.17$}            & $0.94$            \\ \cline{4-6} 
    \multicolumn{1}{|c|}{}                                  & \multicolumn{1}{c|}{}                   & \multicolumn{1}{c|}{}                   & \multicolumn{1}{c|}{Micro}                                                                              & \multicolumn{1}{c|}{$2.34$}            & $1.00$            \\ \hline
    \multicolumn{1}{|c|}{\multirow{2}{*}{Anger}}            & \multicolumn{1}{c|}{\multirow{2}{*}{$2.25$}} & \multicolumn{1}{c|}{\multirow{2}{*}{$1.02$}} & \multicolumn{1}{c|}{Macro}                                                                              & \multicolumn{1}{c|}{$2.13$}            & $0.94$            \\ \cline{4-6} 
    \multicolumn{1}{|c|}{}                                  & \multicolumn{1}{c|}{}                   & \multicolumn{1}{c|}{}                   & \multicolumn{1}{c|}{Micro}                                                                              & \multicolumn{1}{c|}{$2.36$}            & $1.08$            \\ \hline
    \multicolumn{1}{|c|}{\multirow{2}{*}{Surprise}}         & \multicolumn{1}{c|}{\multirow{2}{*}{$2.89$}} & \multicolumn{1}{c|}{\multirow{2}{*}{$1.20$}} & \multicolumn{1}{c|}{Macro}                                                                              & \multicolumn{1}{c|}{$3.69$}            & $0.79$            \\ \cline{4-6} 
    \multicolumn{1}{|c|}{}                                  & \multicolumn{1}{c|}{}                   & \multicolumn{1}{c|}{}                   & \multicolumn{1}{c|}{Micro}                                                                              & \multicolumn{1}{c|}{$2.08$}            & $0.99$            \\ \hline
    \multicolumn{1}{|c|}{\multirow{2}{*}{Fear}}             & \multicolumn{1}{c|}{\multirow{2}{*}{{\ul $2.06$}}} & \multicolumn{1}{c|}{\multirow{2}{*}{$0.92$}} & \multicolumn{1}{c|}{Macro}                                                                              & \multicolumn{1}{c|}{$2.20$}            & $0.89$            \\ \cline{4-6} 
    \multicolumn{1}{|c|}{}                                  & \multicolumn{1}{c|}{}                   & \multicolumn{1}{c|}{}                   & \multicolumn{1}{c|}{Micro}                                                                              & \multicolumn{1}{c|}{{\ul $1.91$}}            & $0.93$            \\ \hline
    \multicolumn{1}{|c|}{\multirow{2}{*}{Disgust}}          & \multicolumn{1}{c|}{\multirow{2}{*}{\textbf{2.95}}} & \multicolumn{1}{c|}{\multirow{2}{*}{$1.13$}} & \multicolumn{1}{c|}{Macro}                                                                              & \multicolumn{1}{c|}{$3.46$}            & $0.92$            \\ \cline{4-6} 
    \multicolumn{1}{|c|}{}                                  & \multicolumn{1}{c|}{}                   & \multicolumn{1}{c|}{}                   & \multicolumn{1}{c|}{Micro}                                                                              & \multicolumn{1}{c|}{$2.45$}            & $1.09$            \\ \hline
    \end{tabular}
    \caption{Table of averages and standard deviations referring to the Emotion Intensity score, which is related to the answers to $Q2$ in Table~\ref{tab:questions-p1}. The lowest averages are highlighted in underline, and the highest averages highlighted in bold, both for emotions in general and for emotions divided by macro and macro expressions.
    }
    \label{tab:emotion-intensity-avg}
\end{table}

We found significant results in the main effects of Video Emotion (\textit{F}(5, 870)=$25.86$ and \textit{p}$<.001$; no equal variance using the Bartlett test - Stats=$21.86$ and \textit{p}$<.001$) and Type of Expression (\textit{F}(1, 874)=$117.15$ and \textit{p}$<.001$), and also in the interaction (\textit{F}(11, 864)=$29.40$ and \textit{p}$<.001$; equal variance - Stats=$11.67$ and \textit{p}$=.38$) between these two variables. Firstly, in this section, we used the post hoc Tamhane test for no equal variance and the Tukey HSD test for equal variance. Regarding the results of the post hoc test (\textit{p}-values are presented in Table~\ref{tab:anova-intensity}), as we can see in the averages (averages are presented in Table~\ref{tab:emotion-intensity-avg}) of the Emotion Intensity score, participants evaluated happiness, surprise, and disgust as the most intense emotions. Statistically, the three emotions had similar intensity averages and were higher than the others. Furthermore, participants rated fear, sadness, and anger as statistically less intense emotions than happiness, surprise, and disgust, and similarly intense to each other. 

Regarding the Type of Expression, videos containing macro expressions (AVG=$2.90$; SD=$1.17$) were considered more intense than videos containing micro expressions (AVG=$2.20$; SD=$1.01$). About the interaction between the variables Video Emotion and Type of Expression, firstly, evaluating the post hoc results on the comparisons involving macro expressions (averages presented in Table~\ref{tab:emotion-intensity-avg} and \textit{p}-values presented in Table~\ref{tab:anova-intensity-interaction}), we can see that people evaluated happiness, surprise and disgust as macro expressions being more intense than the other macro emotions. Regarding the results of the micro expressions, we found only one significant result between the paired comparisons of the micro expression of disgust and fear. As we can see in Table~\ref{tab:emotion-intensity-avg}, the micro expression of disgust had the highest average intensity, and the fear expression had the lowest mean intensity. In the comparisons between the macro versus micro expressions, we only had significant results (Table~\ref{tab:anova-intensity-interaction}) when they were the emotions happiness, disgust, and surprise, being the macro expressions versions of the emotions were considered more intense than their micro expressions versions. In other words, people perceived more intensity in the macro expressions than in the micro expressions of happiness, disgust, and surprise. \textbf{Consequently, we reject $H0_1$ for driven animations once different emotions and the type of expression (micro and macro) make people evaluate the intensities of the emotions differently. In general, macro expressions were considered more intense than micro expressions.}

\subsection{Comparisons on Perceived Comfort for Driven Animation - $H0_2$}
\label{sec:comfortResults}

As in Section~\ref{sec:intensityResults}, about $H0_2$, we used a score based on the 5 Likert scales to analyze the results of perceived comfort presented by question $Q3$ in Table~\ref{tab:questions-p1}, with 1 = ``Very uncomfortable" and 5 = ``Very comfortable". In all analyses in this section, we use the Perceived Comfort Scale as the response variable and the Video Emotion and Type of Expression (Micro or Macro) as predictor variables. The significance level used was 5\%.

\begin{table}[!htb]
    \centering
    \footnotesize
    \begin{tabular}{|cccccc|}
    \hline
    \multicolumn{6}{|c|}{\textbf{Perceived Comfort}}                                                                                                                                                                                                                                                       \\ \hline
    \multicolumn{1}{|c|}{\multirow{2}{*}{\textbf{Emotion}}} & \multicolumn{2}{c|}{\textbf{User’s Answer}}                                       & \multicolumn{1}{c|}{\multirow{2}{*}{\textbf{\begin{tabular}[c]{@{}c@{}}Macro / \\ Micro\end{tabular}}}} & \multicolumn{2}{c|}{\textbf{User’s Answer}}      \\ \cline{2-3} \cline{5-6} 
    \multicolumn{1}{|c|}{}                                  & \multicolumn{1}{c|}{\textbf{Avg}}       & \multicolumn{1}{c|}{\textbf{STD}}       & \multicolumn{1}{c|}{}                                                                                   & \multicolumn{1}{c|}{\textbf{Avg}} & \textbf{STD} \\ \hline
    \multicolumn{1}{|c|}{\multirow{2}{*}{Happiness}}        & \multicolumn{1}{c|}{\multirow{2}{*}{\textbf{3.34}}} & \multicolumn{1}{c|}{\multirow{2}{*}{$0.88$}} & \multicolumn{1}{c|}{Macro}                                                                              & \multicolumn{1}{c|}{$3.35$}            & $0.97$            \\ \cline{4-6} 
    \multicolumn{1}{|c|}{}                                  & \multicolumn{1}{c|}{}                   & \multicolumn{1}{c|}{}                   & \multicolumn{1}{c|}{Micro}                                                                              & \multicolumn{1}{c|}{$3.32$}            & $0.78$            \\ \hline
    \multicolumn{1}{|c|}{\multirow{2}{*}{Sadness}}          & \multicolumn{1}{c|}{\multirow{2}{*}{{\ul $3.04$}}} & \multicolumn{1}{c|}{\multirow{2}{*}{$0.81$}} & \multicolumn{1}{c|}{Macro}                                                                              & \multicolumn{1}{c|}{$3.15$}            & $0.89$            \\ \cline{4-6} 
    \multicolumn{1}{|c|}{}                                  & \multicolumn{1}{c|}{}                   & \multicolumn{1}{c|}{}                   & \multicolumn{1}{c|}{Micro}                                                                              & \multicolumn{1}{c|}{$2.93$}            & $0.71$            \\ \hline
    \multicolumn{1}{|c|}{\multirow{2}{*}{Anger}}            & \multicolumn{1}{c|}{\multirow{2}{*}{$3.06$}} & \multicolumn{1}{c|}{\multirow{2}{*}{$0.85$}} & \multicolumn{1}{c|}{Macro}                                                                              & \multicolumn{1}{c|}{{\ul $2.91$}}            & $0.89$            \\ \cline{4-6} 
    \multicolumn{1}{|c|}{}                                  & \multicolumn{1}{c|}{}                   & \multicolumn{1}{c|}{}                   & \multicolumn{1}{c|}{Micro}                                                                              & \multicolumn{1}{c|}{$3.21$}            & $0.78$            \\ \hline
    \multicolumn{1}{|c|}{\multirow{2}{*}{Surprise}}         & \multicolumn{1}{c|}{\multirow{2}{*}{$3.30$}} & \multicolumn{1}{c|}{\multirow{2}{*}{$0.77$}} & \multicolumn{1}{c|}{Macro}                                                                              & \multicolumn{1}{c|}{\textbf{3.43}}            & $0.79$            \\ \cline{4-6} 
    \multicolumn{1}{|c|}{}                                  & \multicolumn{1}{c|}{}                   & \multicolumn{1}{c|}{}                   & \multicolumn{1}{c|}{Micro}                                                                              & \multicolumn{1}{c|}{$3.16$}            & $0.72$            \\ \hline
    \multicolumn{1}{|c|}{\multirow{2}{*}{Fear}}             & \multicolumn{1}{c|}{\multirow{2}{*}{$3.25$}} & \multicolumn{1}{c|}{\multirow{2}{*}{$0.80$}} & \multicolumn{1}{c|}{Macro}                                                                              & \multicolumn{1}{c|}{$3.20$}            & $0.86$            \\ \cline{4-6} 
    \multicolumn{1}{|c|}{}                                  & \multicolumn{1}{c|}{}                   & \multicolumn{1}{c|}{}                   & \multicolumn{1}{c|}{Micro}                                                                              & \multicolumn{1}{c|}{$3.30$}            & $0.73$            \\ \hline
    \multicolumn{1}{|c|}{\multirow{2}{*}{Disgust}}          & \multicolumn{1}{c|}{\multirow{2}{*}{$3.14$}} & \multicolumn{1}{c|}{\multirow{2}{*}{$0.79$}} & \multicolumn{1}{c|}{Macro}                                                                              & \multicolumn{1}{c|}{$2.97$}            & $0.86$            \\ \cline{4-6} 
    \multicolumn{1}{|c|}{}                                  & \multicolumn{1}{c|}{}                   & \multicolumn{1}{c|}{}                   & \multicolumn{1}{c|}{Micro}                                                                              & \multicolumn{1}{c|}{$3.31$}            & $0.68$            \\ \hline
    \end{tabular}
    \caption{Table of averages and standard deviations referring to the Perceived Comfort score, which is related to the answers to $Q3$ in Table~\ref{tab:questions-p1}. The lowest averages are highlighted (underlined), and the highest averages are highlighted in bold, both for emotions in general and for emotions divided by macro and macro expressions.
    \label{tab:comfort-avg}
    }
\end{table}

We found significant results in the main effect of Video Emotion (\textit{F}(5, 870)=$3.46$ and \textit{p}$=.004$; with equal variance using the Bartlett test - Stats=$3.40$ and \textit{p}$=.63$), and also in the interaction effect (\textit{F}(11, 864)=$3.66$ and \textit{p}$=.002$; with equal variance using the Bartlett test - Stats=$18.66$ and \textit{p}$=.06$) between Video Emotion and Type of Expression. Firstly, as all significant results had equal variance, we used the post hoc Tukey HSD test in this section. Furthermore, we did not find a significant result in the main effect of Type of Expression, that is, people felt similarly comfortable with both micro (AVG=$3.21$; SD=$0.74$) and macro (AVG=$3.17$; SD=$0.89$;) expressions. Regarding the main effect of Video Emotion, we found only one significant result (\textit{p}=$.02$) in comparing the emotions of happiness and sadness. In this case, as we can see in Table~\ref{tab:comfort-avg} regarding the Perceived Comfort averages, people felt more comfortable with the VH expressing happiness than sadness. Regarding the interaction between Video Emotion and Type of Expression, we found only two significant results in the paired comparisons between the macro expression of surprise versus the macro expression of disgust ($.02$) and between the macro expression of surprise versus the macro expression of anger ($.006$). So, \textbf{we cannot reject $H0_2$ since people felt comfortable with micro and macro facial expressions in a similar way. We only found significant differences in the following: \textit{i)} In general, people felt more comfortable with facial expressions of happiness than sadness; and \textit{ii)} In macro expressions, people felt more comfortable with surprise than disgust and anger.}

\subsection{Comparisons between Macro and Micro Emotions Recognition in Real Actors and VHs videos - $H0_3$}
\label{sec:comparisonVideosResults}

To analyze the answers to question $Q1$ (2AFC) presented in Table~\ref{tab:questions-p2} and related to the experiment part 2 ($H0_3$), we used the \textit{Chi-squared Goodness-of-fit} test, which can compare the percentages of correct answers in $Q1$ in relation to the main variables Video Emotion and Type of Expression, and the interaction between them. The significance level used was 5\%.

However, we did not find any significant results, meaning the hypothesis cannot be rejected. In other words, both for Video Emotion (X$^2$(1, 290)=$1.86$ and \textit{p} = $.17$) and Expression Type (X$^2$(1, 290)=$1.86$ and \textit{p} = $.17$), and also for the interaction between them (X$^2$(3, 288)=$5.93$ and \textit{p} = $.11$), the percentages of correct answers to $Q1$ were similar. For example, regarding Video Emotion, as we can see in the percentages column of Table~\ref{tab:percentage-answers-p2}, the percentage of correct answers for videos in which VHs expressed happiness was higher than for videos containing anger, but without statistically significant difference. Regarding Type of Expression, micro expression had a higher percentage than macro expression, but again, it did not have a significant result. Regarding the interaction between Type of Expression and Video Emotion, as we can see in Table~\ref{tab:percentage-answers-p2}, the video with VH expressing a macro expression of anger had the lowest percentage. However, the statistical test result showed similarities between the last four percentages of Table~\ref{tab:percentage-answers-p2}: macro happiness, macro anger, micro happiness, and micro anger. Furthermore, all percentages of correct answers were greater than 60\%, and the general percentage was above 89\%. In other words, almost 90\% of participants responded following the video reference. \textbf{So, we cannot reject $H0_3$ since people correctly identified, for both micro and macro expressions, which
videos involving VHs were related to videos of real actors.}

\begin{table}[!htb]
    \centering
    \begin{tabular}{|c|c|}
    \hline
                      & Percentage (\%) \\ \hline
    General           & 89.04\%      \\ \hline
    Macro             & 81.50\%      \\ \hline
    Micro             & 96.57\%      \\ \hline
    Happiness         & 96.57\%      \\ \hline
    Anger             & 81.50\%      \\ \hline
    Macro - Happiness & \textbf{97.26}\%      \\ \hline
    Macro - Anger     & {\ul 65.75\%}      \\ \hline
    Micro - Happiness & 95.89\%      \\ \hline
    Micro - Anger     & \textbf{97.26}\%      \\ \hline
    \end{tabular}
    \caption{Percentages of correct answers (2AFC) to question $Q1$, which was presented in Table~\ref{tab:questions-p2}, regarding videos with VHs. The lowest percentage is highlighted (underlined), and the highest is highlighted in bold.
    }
    \label{tab:percentage-answers-p2}
\end{table}

\section{Discussion}
\label{sec:discussion}

This section discusses the results presented in Section~\ref{sec:results} about our hypotheses. About $H0_1$, on average, the recognition of emotions in both macro and micro expressions was less accurate in our work than in the competitive studies \cite{andreotti2021perception}, \cite{queiroz2014investigating}, and \cite{montanha2023revisiting}. Regarding macro expressions, we could not statistically compare due to missing data. Looking at the percentage values of each emotion, the results showed that our work obtained higher percentages of correct answers than other studies when the VHs presented macro expressions of happiness and surprise. Analyzing qualitatively by observing the VH macro expressions in Figure~\ref{fig:macro-expressions}, we were able to see characteristic movements (cited by Ekman~\cite{friesen1983emfacs,EkmanFACS}) of the emotions of happiness and surprise, which may have facilitated the recognition of both. For example, in the emotion of happiness, VH presented a smile, while in the emotion of surprise, VH opened its mouth. So, in these two cases, we can say that the macro expressions were correctly transferred from live actors to VHs. \textcolor{black}{On the other hand, when people had difficulty recognizing the macro expressions of fear, sad and disgust, one can observe that the videos do not present the characteristics proposed by Ekman's Facial Action Coding System~\footnote{https://imotions.com/blog/learning/research-fundamentals/facial-action-coding-system/}.}

However, concerning the micro expressions, statistically, the participants in our work recognized the expressions less correctly than those in the studies by Montanha et al. and Queiroz et al., and similarly compared to the work by Andreotti et al. The micro expressions in our work which present the highest percentages of correct answers were happiness and disgust (both had percentages above 50\%). Analyzing qualitatively by looking at Figure~\ref{fig:micro-expressions}, we can see that VH presented a slight smile in happiness and lowered its eyebrow in disgust. \textcolor{black}{Furthermore, the micro expressions of anger and surprise had slightly different results than those of macro expressions. Regarding anger, VH showed an eyebrow-raising movement, which may explain the higher recognition percentage compared with the macro expression. The micro expression of surprise did not present the characteristic mouth-opening movement, which may explain the difficulty people have in recognizing this micro expression.} 

Therefore, for macro expressions, we obtained less than 50\% correct answers on emotion recognition and less than 30\% for micro expressions. So, overall, we can conclude that micro and macro expressions can be recognized more easily in animations modeled by professionals such as designers, artists, and animators than through driven facial animations. It can be explained by the fact that designers can improve and evaluate the animation a posteriori, while the transferring process using Computer Vision techniques, in general, provides a smoothing effect in the facial AUs. Specifically concerning micro expressions, our work involving a realistic VH had similar results to Andreotti et al.'s work, represented by a cartoon VH, i.e., similar perception even in different realism of VHs. This should be further investigated in future work. Still regarding the micro expressions, Tastemirova et al.~\cite{tastemirova2022microexpressions} stated that solutions based on metrics that use Artificial Intelligence tend not to generate micro expressions adequately or to distort the general appearance of the VH. 
Finally, our results can suggest to the industry that emotions can be better recognized if animations are artist-based.

Still, regarding $H0_1$, if the industry still prefers to use performance-driven facial animations related to macro expressions, the most accurately recognized emotions were happiness and surprise, while the least recognized was anger. Of the micro expressions, happiness, anger, and disgust were the most recognized emotions, and fear was the least recognized. Comparing macro versus micro, people recognized the emotions of happiness, sadness, surprise and fear more accurately in the macro expressions versions than in the micro expressions, while disgust and anger were more recognized in the micro expressions than in the macro ones. 

Regarding perceived comfort ($H0_2$), happiness was the emotion with the highest average perceived comfort, while surprise macro expression obtained the highest average perceived comfort. This is in line with the studies of Tinwell et al.~\cite{tinwell2011facial,tinwell2011effect}, where the surprise was one of the facial emotions that conveyed the least uncanny sensation to the participants. Furthermore, anger was considered one of the macro emotions that conveyed less comfort, which is in the same line as the work of Zell et al.~\cite{zell2015stylize}, whose results showed that participants considered anger expressed by VHs to have less appeal than other emotions. Our result was also in line with the work of M{\"a}k{\"a}r{\"a}inen et al.~\cite{makarainen2014exaggerating}, whose results showed that an animated face with an exaggerated surprise emotion caused less strangeness than expected. Overall, people felt similarly comfortable between macro and micro expressions. As in most cases, the average perceived comfort was greater than 3 (which was the score for neutral comfort on the Likert scale of 5), we can hypothesize that animations of micro and macro facial expressions directed in a realistic VH can exceed the Uncanny Valley~\cite{mori2012uncanny}. Following some studies on the Uncanny Valley, photorealistic VHs~\cite{katsyri2015review,araujo2021perceived,macdorman2016reducing}, especially when modeled using the Metahuman system~\cite{higgins2021ascending}, can overcome the valley, and convey comfort (or not convey feelings of strangeness). So, following our results, both micro and macro expressions conveyed comfort to participants. This could be a positive indication that the industry should use facially driven animations regarding micro and macro expressions.

Regarding $H0_3$, we showed that participants could identify which videos involving VHs with driven animations were related to videos of real people, both about micro and macro expressions.

Another aspect relevant to this paper's discussion regards the transference from actors' videos to VHs. Firstly, in Section~\ref{sec:stimuli}, we showed that, through OpenFace, the intensities of macro and micro expression AUs presented in videos involving real people significantly differed from the transferred videos to CG. Although there is a quantitative difference, people correctly associated real and CG videos in this work. So, we can say that Metahuman's realistic models and its driven animation technology (Live Link Face) can convey emotions from videos of real people, although the intensities of facial expressions are smoothed. For the industry, from small to large companies, it is enough to have computers that run Metahuman+Unreal and smartphones compatible with Live Link Face, which can have photorealistic VHs that express the emotions of real people. Furthermore, this can be important for companies that are unable to have large teams of animators, as anyone can do the process to create animations of micro and macro facial expressions based on real people.

\section{Final Considerations}
\label{sec:final_considerations}

In this work, we carried out a perceptual study to observe whether the macro and micro facial expressions of a VH animated from real actors' motions were perceived similarly to previous studies when the facial expressions were modeled in animation. In addition, we evaluated the performance driven animation transferred from real actors to VHs. Therefore, based on the results of our hypotheses, we raised the following final considerations:

\begin{itemize}
    \item $H0_1$ (People recognize and evaluate intensities of macro and micro emotional expressions with less accuracy in our work). Our work showed trends that modeled animations of micro and macro expressions can make people recognize emotions more accurately than driven animations of micro and macro expressions. 
    \item $H0_2$ (People feel similarly comfortable with macro and micro emotion expressions in driven facial animations) - Our work showed no differences in the perception of comfort between driven animated micro and macro expressions. In general, people tended to feel comfortable with both micro and macro expressions.
    \item $H0_3$ (People can identify macro and micro emotion expressions created through driven facial animation performed by people in real videos) - Transfers of micro and macro facial expressions to VHs were identified with more than 90\% in most cases. In other words, the driven animated micro and macro expressions managed to convey the emotions presented to the participants in our study.
\end{itemize}

\textcolor{black}{This study is important for advancing research involving facial expressions in Psychology and Computer Science. Furthermore, the results of our study also help to advance the level of realism of facial expressions in games and movies. For example, several games and movies have used facial expressions animated from real actors' motions and modeled animations by designers, such as Death Stranding\footnote{https://gamerant.com/games-best-facial-animations/} and Cyberpunk 2077\footnote{https://www.gameanim.com/2020/12/05/cyberpunk-2077-procedural-facial-animation/} and the movie Avatar 2: The Way of Water\footnote{https://www.fxguide.com/fxfeatured/exclusive-joe-letteri-discusses-weta-fxs-new-facial-pipeline-on-avatar-2/}.
} 

Our work had limitations in some processes, such as the subjective choices of videos from the datasets, which could generate different results if other videos were chosen. However, as we mentioned in Section~\ref{sec:stimuli}, we chose the videos representing the most intense micro and macro expressions. In future work, we can use VHs that present micro and macro expressions with different intensity levels (similar to the work of Mcdonnell et al.~\cite{mcdonnell2021model}). Furthermore, we intend to develop the same study approach but aim to measure differences in gender and skin color, among others.

\section*{Acknowledgments}
This study was partly financed by the Coordenação de Aperfeiçoamento de Pessoal de Nivel Superior – Brazil (CAPES) – Finance Code
001; by the Conselho Nacional de Desenvolvimento Científico e Tecnológico - Brazil (CNPq) - Process Numbers 309228/2021-2; 406463/
2022-0; 153641/2024-0; and by the Fundação de Amparo à Pesquisa
do Estado do Rio Grande do Sul (FAPERGS).

\appendix

\FloatBarrier
\section{Statistical Results Tables of Emotion Intensity}
\label{sec:apendIntensity}

In this section, we present the Tables~\ref{tab:anova-intensity} and~\ref{tab:anova-intensity-interaction}, that were related to the Emotion Intensity ANOVA analysis in Section~\ref{sec:intensityResults}. The Table~\ref{tab:anova-intensity} presents the post hoc results of the main effect of Video Emotions (happiness, sadness, anger, fear, disgust and surprise). The Table~\ref{tab:anova-intensity-interaction} presents the post hoc results of the interaction effect between Video Emotions and Type of Expression (macro and micro).

\FloatBarrier

\begin{table}[]
    \centering
    \footnotesize
    \begin{adjustbox}{max width=0.98\linewidth}
    \begin{tabular}{|c|c|c|c|c|c|c|}
    \hline
             & Happiness & Sadness & Anger & Fear & Surprise & Disgust \\ \hline
    Happiness    & $1.00$     & $<.001$*   & $<.001$*     & $<.001$*    & $1.00$        & $.99$       \\ \hline
    Sadness      & $<.001$*     & $1.00$   & $1.00$     & $.37$    & $<.001$*        & $<.001$*       \\ \hline
    Anger    & $<.001$*     & $1.00$   & $1.00$     & $.44$    & $<.001$*        & $<.001$*       \\ \hline
    Fear     & $<.001$*     &  $.37$   & $.44$     & $1.00$    & $<.001$*        & $<.001$*       \\ \hline
    Surprise & $1.00$     & $<.001$*   & $<.001$*     & $<.001$*    & $1.00$        & $.99$       \\ \hline
    Disgust  & $.99$     & $<.001$*   & $<.001$*     & $<.001$*    & $.99$        & $1.00$       \\ \hline
    \end{tabular}
    \end{adjustbox}
    \caption{Confusion matrix of \textit{p}-values related to the post hoc of Video Emotions main effect on the response variable Emotion Intensity score. Significant \textit{p}-values are highlighted with an *, taking into account a significance level of 5\%.}
    \label{tab:anova-intensity}
\end{table}

\begin{table*}[]
    \centering
    \footnotesize
    \begin{adjustbox}{max width=0.98\linewidth}
    \begin{tabular}{|c|cccccc|}
    \hline
    \multirow{2}{*}{Macro}  & \multicolumn{6}{c|}{\textbf{vs. Macro}}                                                                                                                                                                                    \\ \cline{2-7} 
                       & \multicolumn{1}{c|}{\textbf{Happiness}} & \multicolumn{1}{c|}{\textbf{Sadness}} & \multicolumn{1}{c|}{\textbf{Anger}} & \multicolumn{1}{c|}{\textbf{Surprise}} & \multicolumn{1}{c|}{\textbf{Fear}} & \textbf{Disgust} \\ \hline
    \textbf{Happiness} & \multicolumn{1}{c|}{$1.00$}                   & \multicolumn{1}{c|}{$<.001$*}                 & \multicolumn{1}{c|}{$<.001$*}               & \multicolumn{1}{c|}{$1.00$}                  & \multicolumn{1}{c|}{$<.001$*}              &   $.57$               \\ \hline
    \textbf{Sadness}   & \multicolumn{1}{c|}{$<.001$*}                   & \multicolumn{1}{c|}{$1.00$}                 & \multicolumn{1}{c|}{$1.00$}               & \multicolumn{1}{c|}{$<.001$*}                  & \multicolumn{1}{c|}{$1.00$}              &           $<.001$*       \\ \hline
    \textbf{Anger}     & \multicolumn{1}{c|}{$<.001$*}                   & \multicolumn{1}{c|}{$1.00$}                 & \multicolumn{1}{c|}{$1.00$}               & \multicolumn{1}{c|}{$<.001$*}                  & \multicolumn{1}{c|}{$1.00$}              &    $<.001$*              \\ \hline
    \textbf{Surprise}  & \multicolumn{1}{c|}{$1.00$}                   & \multicolumn{1}{c|}{$<.001$*}                 & \multicolumn{1}{c|}{$<.001$*}               & \multicolumn{1}{c|}{$1.00$}                  & \multicolumn{1}{c|}{$<.001$*}              &      $.73$            \\ \hline
    \textbf{Fear}      & \multicolumn{1}{c|}{$<.001$*}                   & \multicolumn{1}{c|}{$1.00$}                 & \multicolumn{1}{c|}{$1.00$}               & \multicolumn{1}{c|}{$<.001$*}                  & \multicolumn{1}{c|}{$1.00$}              &        $<.001$*          \\ \hline
    \textbf{Disgust}   & \multicolumn{1}{c|}{$.57$}                   & \multicolumn{1}{c|}{$<.001$*}                 & \multicolumn{1}{c|}{$<.001$*}               & \multicolumn{1}{c|}{$.73$}                  & \multicolumn{1}{c|}{$<.001$*}              &           $1.00$       \\ \hline
    \end{tabular}
    \end{adjustbox}

    \begin{adjustbox}{max width=0.98\linewidth}
    \begin{tabular}{|c|cccccc|}
    \hline
    \multirow{2}{*}{Micro}  & \multicolumn{6}{c|}{\textbf{ vs. Micro}}                                                                                                                                                                                    \\ \cline{2-7} 
                       & \multicolumn{1}{c|}{\textbf{Happiness}} & \multicolumn{1}{c|}{\textbf{Sadness}} & \multicolumn{1}{c|}{\textbf{Anger}} & \multicolumn{1}{c|}{\textbf{Surprise}} & \multicolumn{1}{c|}{\textbf{Fear}} & \textbf{Disgust} \\ \hline
    \textbf{Happiness} & \multicolumn{1}{c|}{$1.00$}                   & \multicolumn{1}{c|}{$.58$}                 & \multicolumn{1}{c|}{$.52$}               & \multicolumn{1}{c|}{$1.00$}                  & \multicolumn{1}{c|}{$.99$}              &     $.19$             \\ \hline
    \textbf{Sadness}   & \multicolumn{1}{c|}{$.58$}                   & \multicolumn{1}{c|}{$1.00$}                 & \multicolumn{1}{c|}{$1.00$}               & \multicolumn{1}{c|}{$.77$}                  & \multicolumn{1}{c|}{$.10$}              &         $1.00$         \\ \hline
    \textbf{Anger}     & \multicolumn{1}{c|}{$.51$}                   & \multicolumn{1}{c|}{$1.00$}                 & \multicolumn{1}{c|}{$1.00$}               & \multicolumn{1}{c|}{$.70$}                  & \multicolumn{1}{c|}{$.09$}              &        $1.00$          \\ \hline
    \textbf{Surprise}  & \multicolumn{1}{c|}{$1.00$}                   & \multicolumn{1}{c|}{$.77$}                 & \multicolumn{1}{c|}{$.70$}               & \multicolumn{1}{c|}{$1.00$}                  & \multicolumn{1}{c|}{$.98$}              &       $.34$           \\ \hline
    \textbf{Fear}      & \multicolumn{1}{c|}{$.99$}                   & \multicolumn{1}{c|}{$.10$}                 & \multicolumn{1}{c|}{$.09$}               & \multicolumn{1}{c|}{$.98$}                  & \multicolumn{1}{c|}{$1.00$}              &        $.02$*          \\ \hline
    \textbf{Disgust}   & \multicolumn{1}{c|}{$.19$}                   & \multicolumn{1}{c|}{$1.00$}                 & \multicolumn{1}{c|}{$1.00$}               & \multicolumn{1}{c|}{$.34$}                  & \multicolumn{1}{c|}{$.02$*}              &           $1.00$       \\ \hline
    \end{tabular}
    \end{adjustbox}

    \begin{tabular}{|c|cccccc|}
    \hline
    \multirow{2}{*}{Macro}  & \multicolumn{6}{c|}{\textbf{ vs. Micro}}                                                                                                                                                                                    \\ \cline{2-7} 
                       & \multicolumn{1}{c|}{\textbf{Happiness}} & \multicolumn{1}{c|}{\textbf{Sadness}} & \multicolumn{1}{c|}{\textbf{Anger}} & \multicolumn{1}{c|}{\textbf{Surprise}} & \multicolumn{1}{c|}{\textbf{Fear}} & \textbf{Disgust} \\ \hline
    \textbf{Happiness} & \multicolumn{1}{c|}{$<.001$*}                   & \multicolumn{1}{c|}{$<.001$*}                 & \multicolumn{1}{c|}{$<.001$*}               & \multicolumn{1}{c|}{$<.001$*}                  & \multicolumn{1}{c|}{$<.001$*}              &   $<.001$*               \\ \hline
    \textbf{Sadness}   & \multicolumn{1}{c|}{$.99$}                   & \multicolumn{1}{c|}{$.98$}                 & \multicolumn{1}{c|}{$.97$}               & \multicolumn{1}{c|}{$1.00$}                  & \multicolumn{1}{c|}{$.70$}              &       $.74$           \\ \hline
    \textbf{Anger}     & \multicolumn{1}{c|}{$1.00$}                   & \multicolumn{1}{c|}{$.93$}                 & \multicolumn{1}{c|}{$.89$}               & \multicolumn{1}{c|}{$1.00$}                  & \multicolumn{1}{c|}{$.88$}              &       $.55$           \\ \hline
    \textbf{Surprise}  & \multicolumn{1}{c|}{$<.001$*}                   & \multicolumn{1}{c|}{$<.001$*}                 & \multicolumn{1}{c|}{$<.001$*}               & \multicolumn{1}{c|}{$<.001$*}                  & \multicolumn{1}{c|}{$<.001$*}              &    $<.001$*              \\ \hline
    \textbf{Fear}      & \multicolumn{1}{c|}{$.98$}                   & \multicolumn{1}{c|}{$.99$}                 & \multicolumn{1}{c|}{$.99$}               & \multicolumn{1}{c|}{$.99$}                  & \multicolumn{1}{c|}{$.52$}              &       $.83$           \\ \hline
    \textbf{Disgust}   & \multicolumn{1}{c|}{$<.001$*}                   & \multicolumn{1}{c|}{$<.001$*}                 & \multicolumn{1}{c|}{$<.001$*}               & \multicolumn{1}{c|}{$<.001$*}                  & \multicolumn{1}{c|}{$<.001$*}              &       $<.001$*           \\ \hline
    \end{tabular}
    \caption{Confusion matrix of \textit{p}-values related to the post hoc of the interaction effect between Video Emotion and Type of Expression, with the Emotion Intensity score as the response variable. Significant \textit{p}-values are highlighted with an *, taking into account a significance level of 5\%.}
    \label{tab:anova-intensity-interaction}
\end{table*}

\FloatBarrier


\bibliographystyle{elsarticle-num} 
\bibliography{cas-refs}

\begin{thebibliography}{10}
\expandafter\ifx\csname url\endcsname\relax
  \def\url#1{\texttt{#1}}\fi
\expandafter\ifx\csname urlprefix\endcsname\relax\def\urlprefix{URL }\fi
\expandafter\ifx\csname href\endcsname\relax
  \def\href#1#2{#2} \def\path#1{#1}\fi

\bibitem{magnenat2003automatic}
N.~Magnenat-Thalmann, H.~Seo, F.~Cordier, Automatic modeling of animatable virtual humans-a survey, in: Fourth International Conference on 3-D Digital Imaging and Modeling, 2003. 3DIM 2003. Proceedings., IEEE, 2003, pp. 2--10.

\bibitem{higgins2021ascending}
D.~Higgins, D.~Egan, R.~Fribourg, B.~Cowan, R.~McDonnell, Ascending from the valley: Can state-of-the-art photorealism avoid the uncanny?, in: ACM Symposium on Applied Perception 2021, 2021, pp. 1--5.

\bibitem{lewis2014practice}
J.~P. Lewis, K.~Anjyo, T.~Rhee, M.~Zhang, F.~H. Pighin, Z.~Deng, Practice and theory of blendshape facial models., Eurographics (State of the Art Reports) 1~(8) (2014) 2.

\bibitem{ekman1969nonverbal}
P.~Ekman, W.~V. Friesen, Nonverbal leakage and clues to deception, Psychiatry 32~(1) (1969) 88--106.

\bibitem{higgins2023investigating}
D.~Higgins, Y.~Zhan, B.~R. Cowan, R.~McDonnell, Investigating the effect of visual realism on empathic responses to emotionally expressive virtual humans, in: ACM Symposium on Applied Perception 2023, 2023, pp. 1--7.

\bibitem{friesen1983emfacs}
W.~V. Friesen, P.~Ekman, et~al., Emfacs-7: Emotional facial action coding system, Unpublished manuscript, University of California at San Francisco 2~(36) (1983) 1.

\bibitem{EkmanFACS}
P.~Ekman, W.~V.~Friesen, Facial action coding system (facs) (01 2002).

\bibitem{Ekman2013AnAF}
P.~Ekman, An argument for basic emotions, 2013.

\bibitem{araujo2021perceived}
V.~Araujo, J.~Melgare, B.~M. Dalmoro, S.~R. Musse, Is the perceived comfort with cg characters increasing with their novelty?, IEEE Computer Graphics and Applications 42~(1) (2021) 32--46.

\bibitem{katsyri2015review}
J.~K{\"a}tsyri, K.~F{\"o}rger, M.~M{\"a}k{\"a}r{\"a}inen, T.~Takala, A review of empirical evidence on different uncanny valley hypotheses: support for perceptual mismatch as one road to the valley of eeriness, Frontiers in psychology 6 (2015) 390.

\bibitem{mori2012uncanny}
M.~Mori, K.~F. MacDorman, N.~Kageki, The uncanny valley [from the field], IEEE Robotics \& automation magazine 19~(2) (2012) 98--100.

\bibitem{tinwell2011facial}
A.~Tinwell, M.~Grimshaw, D.~A. Nabi, A.~Williams, Facial expression of emotion and perception of the uncanny valley in virtual characters, Computers in Human behavior 27~(2) (2011) 741--749.

\bibitem{makarainen2014exaggerating}
M.~M{\"a}k{\"a}r{\"a}inen, J.~K{\"a}tsyri, T.~Takala, Exaggerating facial expressions: A way to intensify emotion or a way to the uncanny valley?, Cognitive Computation 6 (2014) 708--721.

\bibitem{barros2019face}
J.~M.~D. Barros, V.~Golyanik, K.~Varanasi, D.~Stricker, Face it!: A pipeline for real-time performance-driven facial animation, in: 2019 IEEE International Conference on Image Processing (ICIP), IEEE, 2019, pp. 2209--2213.

\bibitem{liu2016real}
S.~Liu, X.~Yang, Z.~Wang, Z.~Xiao, J.~Zhang, Real-time facial expression transfer with single video camera, Computer Animation and Virtual Worlds 27~(3-4) (2016) 301--310.

\bibitem{peres2023can}
V.~M.~X. Peres, G.~P. Dal~Molin, S.~R. Musse, Can we truly transfer an actor's genuine happiness to avatars? an investigation into virtual, real, posed and spontaneous faces, in: Proceedings of the 22nd Brazilian Symposium on Games and Digital Entertainment, 2023, pp. 56--65.

\bibitem{macdorman2016reducing}
K.~F. MacDorman, D.~Chattopadhyay, Reducing consistency in human realism increases the uncanny valley effect; increasing category uncertainty does not, Cognition 146 (2016) 190--205.

\bibitem{araujo2023evaluating}
V.~F. de~Andrade~Araujo, A.~B. Costa, S.~R. Musse, Evaluating the uncanny valley effect in dark colored skin virtual humans, in: 2023 36th SIBGRAPI Conference on Graphics, Patterns and Images (SIBGRAPI), IEEE, 2023, pp. 1--6.

\bibitem{katsyri2019virtual}
J.~K{\"a}tsyri, B.~de~Gelder, T.~Takala, Virtual faces evoke only a weak uncanny valley effect: an empirical investigation with controlled virtual face images, Perception 48~(10) (2019) 968--991.

\bibitem{zell2019perception}
E.~Zell, K.~Zibrek, R.~McDonnell, Perception of virtual characters, in: ACM Siggraph 2019 Courses, 2019, pp. 1--17.

\bibitem{tastemirova2022microexpressions}
A.~Tastemirova, J.~Schneider, L.~C. Kruse, S.~Heinzle, J.~v. Brocke, Microexpressions in digital humans: perceived affect, sincerity, and trustworthiness, Electronic Markets 32~(3) (2022) 1603--1620.

\bibitem{hou2022micro}
T.~Hou, N.~Adamo, N.~Villani, Micro-expressions in animated agents, AHFE International 22 (2022).

\bibitem{queiroz2014investigating}
R.~B. Queiroz, S.~R. Musse, N.~I. Badler, Investigating macroexpressions and microexpressions in computer graphics animated faces, Presence 23~(2) (2014) 191--208.

\bibitem{andreotti2021perception}
L.~Andreotti, M.~L. Weber, T.~L. da~Silva, V.~F. de~Andrade~Araujo, S.~R. Musse, Perception of charisma, comfort, micro and macro expressions in computer graphics characters, in: 2021 20th Brazilian Symposium on Computer Games and Digital Entertainment (SBGames), IEEE, 2021, pp. 107--116.

\bibitem{montanha2023revisiting}
R.~Montanha, G.~Raupp, V.~Gonzalez, Y.~Partichelli, A.~Bins, M.~Ferreira, V.~Araujo, S.~Musse, Revisiting micro and macro expressions in computer graphics characters, in: Proceedings of the 22nd Brazilian Symposium on Games and Digital Entertainment, 2023, pp. 38--45.

\bibitem{zielke2011creating}
M.~Zielke, F.~Dufour, G.~Hardee, R.~Taylor, B.~Jacobs, D.~Blair, A.~Buxkamper, J.~Donahue, B.~Keown, D.~Trinh, Creating micro-expressions and nuanced nonverbal communication in synthetic cultural characters and environments, in: Proceedings of the interservice/industry training, simulation \& education conference (I/ITSEC), 2011.

\bibitem{mavadati2016extended}
M.~Mavadati, P.~Sanger, M.~H. Mahoor, Extended disfa dataset: Investigating posed and spontaneous facial expressions, in: proceedings of the IEEE conference on computer vision and pattern recognition workshops, 2016, pp. 1--8.

\bibitem{davison2016samm}
A.~K. Davison, C.~Lansley, N.~Costen, K.~Tan, M.~H. Yap, Samm: A spontaneous micro-facial movement dataset, IEEE transactions on affective computing 9~(1) (2016) 116--129.

\bibitem{bassili1978facial}
J.~N. Bassili, Facial motion in the perception of faces and of emotional expression., Journal of experimental psychology: human perception and performance 4~(3) (1978) 373.

\bibitem{gottschalk1966micromomentary}
L.~A. Gottschalk, A.~H. Auerbach, E.~A. Haggard, K.~S. Isaacs, Micromomentary facial expressions as indicators of ego mechanisms in psychotherapy, Methods of research in psychotherapy (1966) 154--165.

\bibitem{carrigan2020investigating}
E.~Carrigan, K.~Zibrek, R.~Dahyot, R.~McDonnell, Investigating perceptually based models to predict importance of facial blendshapes, in: Proceedings of the 13th ACM SIGGRAPH Conference on Motion, Interaction and Games, 2020, pp. 1--6.

\bibitem{weise2011realtime}
T.~Weise, S.~Bouaziz, H.~Li, M.~Pauly, Realtime performance-based facial animation, ACM transactions on graphics (TOG) 30~(4) (2011) 1--10.

\bibitem{amos2016openface}
B.~Amos, B.~Ludwiczuk, M.~Satyanarayanan, et~al., Openface: A general-purpose face recognition library with mobile applications, CMU School of Computer Science 6~(2) (2016) 20.

\bibitem{mcdonnell2021model}
R.~McDonnell, K.~Zibrek, E.~Carrigan, R.~Dahyot, Model for predicting perception of facial action unit activation using virtual humans, Computers \& Graphics 100 (2021) 81--92.

\bibitem{bailey2017gender}
J.~D. Bailey, K.~L. Blackmore, Gender and the perception of emotions in avatars, in: Proceedings of the Australasian Computer Science Week Multiconference, 2017, pp. 1--8.

\bibitem{ennis2013emotion}
C.~Ennis, L.~Hoyet, A.~Egges, R.~McDonnell, Emotion capture: Emotionally expressive characters for games, in: Proceedings of motion on games, 2013, pp. 53--60.

\bibitem{hyde2014assessing}
J.~Hyde, E.~J. Carter, S.~Kiesler, J.~K. Hodgins, Assessing naturalness and emotional intensity: a perceptual study of animated facial motion, in: Proceedings of the ACM Symposium on Applied Perception, 2014, pp. 15--22.

\bibitem{hyde2016evaluating}
J.~Hyde, E.~J. Carter, S.~Kiesler, J.~K. Hodgins, Evaluating animated characters: facial motion magnitude influences personality perceptions, ACM Transactions on Applied Perception (TAP) 13~(2) (2016) 1--17.

\bibitem{ochs201518}
M.~Ochs, R.~Niewiadomski, C.~Pelachaud, 18 facial expressions of emotions for virtual characters, in: The Oxford handbook of affective computing, Oxford Univ. Press, 2015, p. 261.

\bibitem{zibrek2013evaluating}
K.~Zibrek, L.~Hoyet, K.~Ruhland, R.~McDonnell, Evaluating the effect of emotion on gender recognition in virtual humans, in: Proceedings of the ACM Symposium on Applied Perception, 2013, pp. 45--49.

\bibitem{bornemann2012can}
B.~Bornemann, P.~Winkielman, E.~van~der Meer, Can you feel what you do not see? using internal feedback to detect briefly presented emotional stimuli, International Journal of Psychophysiology 85~(1) (2012) 116--124.

\bibitem{shen2012effects}
X.-b. Shen, Q.~Wu, X.-l. Fu, Effects of the duration of expressions on the recognition of microexpressions, Journal of Zhejiang University Science B 13 (2012) 221--230.

\bibitem{Pumarola_ijcv2019}
A.~Pumarola, A.~Agudo, A.~Martinez, A.~Sanfeliu, F.~Moreno-Noguer, Ganimation: One-shot anatomically consistent facial animation (2019).

\bibitem{wang2019fewshotvid2vid}
T.-C. Wang, M.-Y. Liu, A.~Tao, G.~Liu, J.~Kautz, B.~Catanzaro, Few-shot video-to-video synthesis, in: Conference on Neural Information Processing Systems (NeurIPS), 2019.

\bibitem{savage2013search}
R.~A. Savage, O.~V. Lipp, B.~M. Craig, S.~I. Becker, G.~Horstmann, In search of the emotional face: anger versus happiness superiority in visual search., Emotion 13~(4) (2013) 758.

\bibitem{baltrusaitis2018openface}
T.~Baltrusaitis, A.~Zadeh, Y.~C. Lim, L.-P. Morency, Openface 2.0: Facial behavior analysis toolkit, in: 2018 13th IEEE international conference on automatic face \& gesture recognition (FG 2018), IEEE, 2018, pp. 59--66.

\bibitem{baltruvsaitis2015cross}
T.~Baltru{\v{s}}aitis, M.~Mahmoud, P.~Robinson, Cross-dataset learning and person-specific normalisation for automatic action unit detection, in: 2015 11th IEEE International Conference and Workshops on Automatic Face and Gesture Recognition (FG), Vol.~6, IEEE, 2015, pp. 1--6.

\bibitem{tinwell2011effect}
A.~Tinwell, M.~Grimshaw, D.~Abdel-Nabi, Effect of emotion and articulation of speech on the uncanny valley in virtual characters, in: Affective Computing and Intelligent Interaction: Fourth International Conference, ACII 2011, Memphis, TN, USA, October 9--12, 2011, Proceedings, Part II, Springer, 2011, pp. 557--566.

\bibitem{zell2015stylize}
E.~Zell, C.~Aliaga, A.~Jarabo, K.~Zibrek, D.~Gutierrez, R.~McDonnell, M.~Botsch, To stylize or not to stylize? the effect of shape and material stylization on the perception of computer-generated faces, ACM Transactions on Graphics (TOG) 34~(6) (2015) 1--12.

\end{thebibliography}





\end{document}